\documentclass[twocolumn,txfonts,showpacs,nofootinbib,superscriptaddress,preprintnumbers,amsmath,amssymb,epsfig,color]{aa}

\usepackage{natbib}
\bibpunct{(}{)}{;}{a}{}{,} 

\usepackage[usenames]{color}
\usepackage{graphicx}

\usepackage{units, subfigure,amssymb,amsmath}
\usepackage[latin1]{inputenc}

\begin{document}
\input epsf
\title{Systematic uncertainties on the cosmic-ray transport parameters}
\subtitle{Is it possible to reconcile B/C data with $\delta=1/3$ or $\delta=1/2$?}
\author{
       D. Maurin\inst{1,2,3}
	\and A. Putze\inst{4,5}
	\and L. Derome\inst{4}
} 

\offprints{D. Maurin, {\tt dmaurin@lpnhe.in2p3.fr}}

\institute{Laboratoire de Physique Nucl\'eaire et des Hautes
	Energies ({\sc lpnhe}),
   Universit\'es Paris VI et Paris VII, CNRS/IN2P3,
	Tour 33, Jussieu, Paris,
	75005, France
   \and
   Dept. of Physics and Astronomy, University of Leicester,
   Leicester, LE17RH, UK
   \and
   Institut d'Astrophysique de Paris ({\sc iap}), UMR7095 CNRS,
   Universit\'e Pierre et Marie Curie, 98 bis bd Arago,
   75014 Paris, France
   \and Laboratoire de Physique Subatomique et de
	Cosmologie ({\sc lpsc}),
	Universit\'e Joseph Fourier Grenoble 1, CNRS/IN2P3, Institut Polytechnique de Grenoble,
	53 avenue des Martyrs,
	Grenoble, 38026, France
   \and The Oskar Klein Centre for Cosmoparticle Physics,
   Department of Physics, Stockholm University,
   AlbaNova, SE-10691 Stockholm, Sweden
}

\date{Received / Accepted}

\abstract
{The B/C ratio is used in cosmic-ray physics to constrain the transport parameters.
However, from the same set of data, the various published values show a puzzling
large scatter of these parameters.
}
{We investigate the impact of using different inputs (gas density and hydrogen
fraction in the Galactic disc, source spectral shape, low-energy dependence of the
diffusion coefficient, and nuclear fragmentation cross-sections) on the best-fit
values of the transport parameters. We quantify the systematics produced when varying
these inputs, and compare them to statistical uncertainties. We discuss the
consequences for the slope of the diffusion coefficient $\delta$.
} 
{The analysis relies on the propagation code USINE interfaced with the Minuit
minimisation routines.
}
{We find the typical systematic uncertainties to be greater than the statistical ones.  The
several published values of $\delta$ (from 0.3 to 0.8) can be recovered when varying the
low-energy shape of the diffusion coefficient and the convective wind strength. 
Models including a convective wind are characterised by $\delta\gtrsim 0.6$, which cannot
be reconciled with the expected theoretical values ($1/3$ and $1/2$). However, from a
statistical point of view ($\chi^2$ analysis), models with both reacceleration and
convection|hence large $\delta$|are favoured. The next favoured models in line yield
$\delta$, which can be accommodated with $1/3$ and $1/2$, but require a strong upturn of the
diffusion coefficient at low energy (and no convection).}
{To date, using the best statistical tools, the transport parameter determination is still
plagued by many unknowns at low energy ($\sim$~GeV/n). To disentangle all these
configurations, measurements of the B/C ratio at TeV/n energies and/or combination with other
secondary-to-primary ratios is necessary.
}


\maketitle


\section{Introduction}
A central question in Galactic cosmic ray (GCR) physics lies in the determination of the
transport parameters. The standard procedure consists in fitting a secondary-to-primary
ratio, e.g. B/C. In general, the parameters derived in different studies provide
inconsistent values, especially for the slope $\delta$ of the diffusion coefficient. 
It is important to understand the origin of these differences. 

The first attempts to get the best-fit value of the transport parameters, as well
as their statistical uncertainties, were carried in \citet{2001ApJ...555..585M}
and \citet{2005JCAP...09..010L}. Recently, we implemented a Markov chain Monte
Carlo (MCMC) to estimate the probability-density function (PDF) of the transport
and source parameters for Galactic cosmic rays
\citep[hereafter Papers~I and II]{2009A&A...497..991P,2009A&A...xxx..xxxP}.
This sound statistical technique applied to current B/C data allowed us to estimate the
statistical uncertainties on various parameters. We found typical uncertainties of
$\sim 10-20 \%$.
Although such precision is expected given the accuracy of the data, one may wonder whether
these results are not dominated by systematic uncertainties from the input ingredients and
assumptions made to do the calculation.

This is an important issue since in our previous studies, the best-fit slope observed in a
diffusion/constant-convection/reacceleration propagation model
\citep{2001ApJ...555..585M,2002A&A...394.1039M}|confirmed by our recent MCMC analysis
(Paper~II)|points to $\delta\approx 0.75-0.85$. This is at variance with the result of the
GALPROP code \citep{1998ApJ...509..212S}, where the best-fit models correspond to smaller
$\delta\approx 0.3-0.4$ \citep{2005JCAP...09..010L} in either a diffusion/linear-convection or
a diffusion/reacceleration model. Moreover, none of these results agree with the best-fit
slopes from leaky-box models (with rigidity cut-off), where $\delta\approx 0.5-0.6$ (e.g.,
\citealt{2003ApJS..144..153W}|no reacceleration|, or Paper~I|with reacceleration). Such a
scatter in $\delta$ was also observed in \citet{2001ApJ...547..264J}.

The propagation code USINE (Maurin et al., in preparation) provides the GCR fluxes in both
the framework of the leaky-box model (LBM) and the diffusion model (DM). Associated with an
efficient minimisation tool for finding the best-fit parameters of a model (w/wo convection,
w/wo diffusive reacceleration), it allows these differences to be addressed thoroughly. We also
investigate how sensitive these best-fit parameters are to various input ingredients/parameters.

The paper is organised as follows. In Sect.~\ref{sec:Model}, the DM used is briefly recalled.
In Sect.~\ref{sec:Method}, the free parameters of the study, our methodology, and the input
ingredients|the systematic effects of which are studied in this paper|are described. In
Sect.~\ref{sec:gas}, we discuss the consequences of varying the gas characteristics on the transport
parameter determination. In Sect.~\ref{sec:src}, we focus on the effect of the source spectrum
on the best-fit values for $\delta$ in models with or without convection/reacceleration. In
Sect.~\ref{sec:diff}, a similar analysis is carried out for the low-energy shape of the diffusion
coefficient. We repeat again the analysis in Sect.~\ref{sec:Xsec}, using various production
cross-section sets. In a final step, we explore in Sect.~\ref{sec:biasHEAO} the effect of
biasing B/C HEAO-3 data at high energy.
We summarise, discuss our results, and conclude in Sect.~\ref{sec:conclusion}.

\section{Description of the diffusion model\label{sec:Model}}

The models and the equations are described in Paper~II, to which we refer the reader
for a complete discussion.


\subsection{Diffusion equation}
The differential density $N^j$ of the nucleus $j$ is a function of the total energy $E$ and
the position $\vec{r}$ in the Galaxy.  Assuming steady-state, the transport equation can be
written in a compact form as 
\begin{equation}
{\cal L}^j N^j + \frac{\partial}{\partial E}\left( b^j N^j - c^j \frac{\partial N^j}{\partial E} \right) = {\cal S}^j\;.
\label{eq:CR}
\end{equation}
The operator ${\cal L}$ (we omit the superscript $j$) describes the diffusion
$K(\vec{r},E)$ and convection $\vec{V}(\vec{r})$ in the Galaxy, the decay rate
$\Gamma_{\rm rad}(E)= 1/(\gamma\tau_0)$ for radioactive species, and the destruction
rate $\Gamma_{\rm inel}(\vec{r},E)=\sum_{ISM} n_{\rm ISM}(\vec{r}) v \sigma_{\rm inel}(E)$
on the interstellar matter (ISM):
\begin{equation}
{\cal L}(\vec{r},E) =  -\vec{\nabla} \cdot (K\vec{\nabla}) + \vec{\nabla}\cdot\vec{V} +
     \Gamma_{\rm rad} + \Gamma_{\rm inel}.
\label{eq:operator}
\end{equation}

The coefficients $b$ and $c$ are respectively first and
second order gains/losses in energy, with
\begin{eqnarray}
\label{eq:b}
b\,(\vec{r},E)&=& \big\langle\frac{dE}{dt}\big\rangle_{\rm ion,\,coul.} 
   - \frac{\vec{\nabla}.\vec{V}}{3} E_k\left(\frac{2m+E_k}{m+E_k}\right)
	 \\\nonumber
   &  & + \;\; \frac{(1+\beta^2)}{E} \times K_{pp},\\
\label{eq:c}
c\,(\vec{r},E)&=&  \beta^2 \times K_{pp}.
\end{eqnarray}
The coefficient $K_{pp}$ is the diffusion coefficient in momentum space,
and it can take several forms (see later).

\subsection{Geometry of the Galaxy}
The Galaxy is modelled to be a thin disc of half-thickness $h$, which contains the gas and the
sources of CRs. This disc is embedded in a cylindrical diffusive halo of half-thickness $L$ (where
the gas density is assumed to be equal to 0). A constant wind $\vec{V}(\vec{r})={\rm sign}(z)
\cdot V_c \times \vec{e}_z$, perpendicular to the Galactic plane, is assumed.
In this framework, CRs diffuse in the disc and in the halo independently of their position.
These assumptions allow for semi-analytical solutions of the transport equation, as the
interactions (destruction, spallations, energy gain and losses) are restricted to the thin
disc. Such semi-analytical models reproduce all salient features of full numerical approaches
(e.g., \citealt{1998ApJ...509..212S}).

In this study, the disc half-height is set to $h=100$~pc. It is not a physical parameter {\em
per se} in the thin-disc approximation, but the phenomena occurring in the thin disc are
related to it. The physical parameter is the surface density $\Sigma$: should a different $h$
value be used, a rescaling always allows obtaining the same $\Sigma$.

Considering the radial extension $R$ of the Galaxy to be infinite leads to the 1D version of
the DM. This geometry, used in \citet{2001ApJ...547..264J} and in our Paper~II, is also used
in this analysis. The corresponding sets of equations and their solutions are presented
in the Appendix of Paper~II. They are not repeated here.

\section{Methodology\label{sec:Method}}

As in Papers~I and~II, three different classes of diffusion models are considered. In addition,
for completeness, we also treat the pure diffusion case:
\begin{itemize}  
  \item Model~0~$=\{K_0,\, \delta\}$, i.e. pure diffusion ($V_a = V_c =0$);
  \item Model~I~$=\{K_0,\, \delta\,\, V_c\}$, i.e. no reacceleration ($V_a = 0$);
  \item Model~II~$=\{K_0,\, \delta, \, V_a\}$, i.e. no convection ($V_c = 0$);
  \item Model~III~$=\{K_0,\, \delta, \, V_c, \, V_a\}$.
\end{itemize}
The parameters $K_0$ and $\delta$ come from the standard form assumed for the diffusion
coefficient, namely,
\begin{equation}
K(E)= \beta K_0 {\cal R}^{\delta}.
\label{eq:K(E)}
\end{equation}
%

\subsection{Constrained parameters}

At most, for Model~III, four transport parameters need to be determined from CR data:
\begin{itemize}
   \item $K_0$, the normalisation of the diffusion coefficient (in unit of kpc$^2$~Myr$^{-1}$);
   \item $\delta$, the slope of the diffusion coefficient;
   \item $V_c$, the constant convective wind perpendicular to the disc (km~s$^{-1}$);
   \item $V_a$, the Alfv\'enic speed (km~s$^{-1}$) regulating the reacceleration strength [see Eq.~(\ref{eq:Va})].
\end{itemize}
As in other studies, we use the B/C ratio to constrain these parameters. In DMs, the halo size
of the Galaxy $L$ is an extra free parameter. It cannot be determined solely from the B/C ratio
because of the well-known degeneracy between $K_0$ and $L$. In this study, we choose to fix $L$.
The value of the transport parameters for any other values of $L$ can be obtained from simple
scaling laws, as presented in Fig.~5 of Paper~II (note that  $\delta$ does not depend on
$L$). To keep the discussion as simple as possible, this paper is based on B/C data alone, with
four free parameters $\{K_0,\,\delta,\,V_c,\,V_a\}$ and the halo size set to $L=4$~kpc. 

\subsection{Inputs and default configuration\label{sec:Params_Inputs}}

The inputs that we examine and vary are the following (details and references 
are given in Paper~II, Sect. 2.4):
\begin{itemize}
  \item gas surface density $\Sigma_{\rm ISM}=2hn_{\rm ISM}$ and
	hydrogen fraction $f_{\rm H}$ (in number),
  \item source spectrum $Q^j(E)= q_j \beta^{\eta_S} R^{- \alpha}$ ($q_j$ is scaled to match the
	measured elemental fluxes at $\sim 10$~GeV/n),
  \item spatial/momentum diffusion coefficients using different turbulence models \citep{2002cra..book.....S},
  \item production cross-sections.
\end{itemize}
The reference (default) values for these inputs are gathered in Table~\ref{tab:default}.
\begin{table}[!t]
\caption{Reference values for the inputs.}
\label{tab:default}
\centering
\begin{tabular}{ll} \hline\hline
Input name              &               Default value/dependence/set                  \vspace{0.05cm}\\\hline
& \multicolumn{1}{c}{} \vspace{-0.25cm} \\ 
Gas                     &  $\Sigma=6.17\cdot10^{20}$~cm$^2, \quad f_{\rm H}= 90\%$    \vspace{0.05cm}\\
Source spectrum         &  $\eta_S=-1, \quad \alpha+\delta=2.65$                      \vspace{0.05cm}\\
$K(E)$ and $K_{pp}(E)$  &  Slab Alfv\'en (SA): Eqs.~(\ref{eq:K(E)}) \& (\ref{eq:Va})  \vspace{0.05cm}\\
Cross-sections          &  W03 \citep{2003ApJS..144..153W}                            \vspace{0.05cm}\\
Data                    &  B/C, dataset F$^\dagger$                                   \vspace{0.05cm}\\
\hline
\end{tabular}
{\tiny \\$^\dagger$ 31 data points from IMP7-8, Voyager~1\&2, ACE, HEA0-3, Spacelab, and CREAM04}
\note{\tiny In subsequent tables of the paper, all results obtained with the default inputs are in {\em italics}.}
\end{table}
The default diffusion coefficient $K_{pp}$ (in momentum space) is taken from the model of minimal reacceleration by the
interstellar turbulence \citep{1988SvAL...14..132O,1994ApJ...431..705S}:
\begin{equation}
K_{pp}\times K= \frac{4}{3}\;V_a^2\;\frac{p^2}{\delta\,(4-\delta^2)\,(4-\delta)},
\label{eq:Va}
\end{equation}
where $V_a$ is the Alfv\'enic speed in the medium.

Concerning the experimental data used to fit the B/C ratio, as in Paper~II,
we use the following dataset (denoted F in Paper II), which consists of 31 B/C data points
(see Fig.~\ref{fig:bias_data} of this paper):
IMP7-8~\citep{1987ApJS...64..269G}, Voyager~1\&2 \citep{1999ICRC....3...41L}, 
ACE-CRIS \citep{2006AdSpR..38.1558D}, HEA0-3 \citep{1990A&A...233...96E}, 
Spacelab~\citep{1990ApJ...349..625S},  and CREAM~\citep{2008APh....30..133A}. 


\subsection{Minimisation routine}
In Papers~I and~II, we adapted the MCMC technique for a propagation code. This
allows the PDF of the parameters to be obtained along with their statistical uncertainties.
However, this technique relies on thousands of calculations for a single setting of the inputs,
which is not optimal for speed. Here, we are only interested in the best-fit values, not
in the PDF of the parameters.  Therefore, we use the Minuit library (a CERN library), which
provides minimisation routines. Instead of a few hours of distributed calculations for the
MCMC technique, a few minutes on a workstation are enough to obtain the best-fit
parameters. 

A few configurations have been cross-checked with the MCMC technique, to ensure that the
typical widths of the PDFs remain the same, whatever the input ingredients used in the
calculation. This allows us, for a given propagation model calculated from different input
ingredients, to compare the resulting scatter in the transport parameter values|systematic
uncertainties (hereafter {\em SystUnc})|to the typical statistical uncertainties
(hereafter {\em StatUnc}) found with the MCMC technique.

\section{Influence of the gas description\label{sec:gas}}
We start by varying the gas parameters. This is discussed for the most general class of the
DM, i.e. Model~III (allowing for both convection and reacceleration). For obvious reasons
(see below), our conclusions also hold for Models~0, I, or~II.

In 1D models, the surface density of the gas in the model corresponds to the average of the {\em
true} gas surface density (which depends on the position in the Galaxy) over the effective diffusion
volume \citep{2003A&A...402..971T}. This is a reasonable assumption to make for stable nuclei (see
Paper~II). However, the details of the volume over which to calculate this average are not
straightforward. Moreover, even if a more realistic distribution of gas were to be used,
the latter is not free of uncertainties. We thus allow for some uncertainty in this input.
We also vary the fraction of hydrogen relative to helium.

  \subsection{Influence of the hydrogen fraction}
In Table~\ref{tab:Systematics_gas}, the second and third lines  (compare with the first line)
show the effect of changing the hydrogen fraction in the ISM: the transport parameters are
changed at most by $\sim 5\%$. There is a systematic trend for $\delta$ to decrease with
smaller fractions of helium in the gas. However, the uncertainty on the hydrogen and helium
fraction is not more than a few \%. We therefore conclude that this has no strong impact
on the derived transport parameters.

  \subsection{Influence of the surface density} 
The fourth and fifth lines show the effect of changing the surface gas density $\Sigma_{\rm
ISM}=2hn_{\rm ISM}$. Whereas it has a small impact on the diffusion slope $\delta$, the other
transport parameters are strongly affected. The change in the parameters can be understood if we look at
the grammage of the DM \citep{2002A&A...394.1039M}. In the purely diffusive regime, we have
(e.g., \citealt{2006astro.ph.12714M})
      \begin{equation}
               \label{pure-DM}
               \langle x\rangle^{pure-DM} =\frac{\Sigma_{\rm ISM} \bar{m} vL}{2K}\;,
       \end{equation}
where $\bar{m}$ is the mean mass of the ISM. Let $K^{\rm ref}_0$, $\delta^{\rm ref}$, $V_c^{\rm ref}$,
and $V_a^{\rm ref}$ be the best-fit parameters obtained for a surface density of gas $\Sigma^{\rm ref}$
(first line of the Table). We remind that $L$ is fixed here. If the surface gas density is rescaled,
i.e. $\Sigma^{\rm new}= x\times \Sigma^{\rm ref}$, in order to keep the same grammage in
Eq.~(\ref{pure-DM}), we need to have $K^{\rm new}_0=x\times K_0^{\rm ref}$. This is what we get
in Table~\ref{tab:Systematics_gas}.
\begin{table}[!t]
\caption{Best-fit transport parameters for different ISM.}
\label{tab:Systematics_gas}
\centering
\begin{tabular}{ccccccc} \hline\hline
\multicolumn{2}{c}{Gas} & $\!\!\!K_0^{\rm best}\times 10^2\!$ & $\delta^{\rm best}$ &  $V_c^{\rm best}$  &  $V_a^{\rm best}$  & $\!\!\!\!\chi^2$/d.o.f$\!\!\!\!$   \\
$\Sigma$   & $f_{\rm H}$ & $\!\!\!$(kpc$^2\,$Myr$^{-1}$)$\!\!$ &  &  $\!\!\!\!$(km$\;$s$^{-1}$)$\!\!$  & $\!\!$(km$\;$s$^{-1}$)$\!\!\!\!$   & \\\hline
& \multicolumn{1}{c}{} \vspace{-0.20cm} \\ 
${\it \Sigma^{\rm ref}}$ & ${\it f^{\rm ref}_{\rm H}}$                                     & {\it 0.48} & {\it 0.86} & {\it 18.8} & {\it 38.0} & {\it 1.47}    \vspace{0.15cm}\\  
$\!\!\Sigma^{\rm ref}$ & $95\%$ & 0.53 & 0.83 & 18.6 & 38.1 & 1.24    \vspace{0.00cm}\\  
$\!\!\Sigma^{\rm ref}$ & $80\%$ & 0.41 & 0.90 & 19.3 & 37.7 & 2.14    \vspace{0.15cm}\\  
$\frac{1}{2}\,\Sigma^{\rm ref}$ & $f^{\rm ref}_{\rm H}$                              & 0.25 & 0.85 & 9.5  & 19.4 & 1.45    \vspace{0.00cm}\\  
$2\,\Sigma^{\rm ref}$&$f^{\rm ref}_{\rm H}$                                        & 0.92 & 0.86 & 37.3 & 74.6 & 1.51    \vspace{0.10cm}\\  
\hline
\end{tabular}
\note{\tiny Model III for $L=4$~kpc. For all settings, we keep fixed $\langle
n_{e^-}\rangle=0.033$ and $T_e\sim10^4\unit{K}$. The reference parameters for the surface density
$\Sigma^{\rm ref}$ and the hydrogen fraction $f^{\rm ref}_{\rm H}$ are given
in Table~\ref{tab:default}.\vspace{-0.25cm}}
\end{table}

The same reasoning holds for the convective wind. In presence of a constant wind, the full
expression for the grammage reads (e.g., \citealt{2006astro.ph.12714M})
       \begin{equation}
			 \label{eq:re-grammage_Vc}
               \left<x\right>^{V_c}\equiv
               \frac{\Sigma_{\rm ISM} \bar{m}  v}{2V_c} \left[1-e^{-\frac{V_cL}{K}}\right]\;.
       \end{equation}
With the above rescaling for $\Sigma^{\rm new}$ and $K^{\rm new}_0$, we get $V^{\rm
new}_c=x\times V_c^{\rm ref}$. Again, this is recovered in Table~\ref{tab:Systematics_gas}.

The last parameter is the Alfv\'enic speed, for which we need to consider the whole transport
equation.  The speed $V_a$ is used to define $K_{pp}$ [see Eq.~(\ref{eq:Va})], which appears in
the terms $b$ and $c$ [Eqs.~(\ref{eq:b}) and (\ref{eq:c})] of Eq.~(\ref{eq:CR}). If we consider
the latter equation and the transport operator ${\cal L}$ in Eq.~(\ref{eq:operator}), the rescaling
$\Sigma^{\rm new}=x\times \Sigma^{\rm ref}$ leads to ${\cal L}^{\rm new}= x\times {\cal L}^{\rm
ref}$. This implies $b^{\rm new}= x\times b^{\rm ref}$, $c^{\rm new}= x\times c^{\rm ref}$, and
${\cal S}^{\rm new}= x\times {\cal S}^{\rm ref}$. Let us consider the three terms each in turn.
For $b$, if we look at Eq.~(\ref{eq:b}), this $x$ factor is automatically ensured for the Coulomb
and ionisation losses (they depend on the gas surface density), and also for adiabatic losses (as
$V^{\rm new}_c=x\times V_c^{\rm ref}$). This implies $K_{pp}^{\rm new}= x\times K_{pp}^{\rm ref}$.
From Eq.~(\ref{eq:Va}), we have $K_{pp}\propto V_a^2/K_0$, such that the previous equality yields
\[
   (V_a^{\rm new})^2/K_0^{\rm new} = x \times (V_a^{\rm ref})^2/K_0^{\rm ref}.
\]
This gives $V_a^{\rm new}= x \times V_a^{\rm ref}$ (the same information is
given by the $c$ term), in agreement with the results of Table~\ref{tab:Systematics_gas}.

  \subsection{Summary for the gas surface density}
An uncertainty of $x\%$ on the gas surface density leads to {\em SystUnc} of $x\%$ on $K_0$,
$V_c$ and $V_a$, but it leaves the diffusion slope $\delta$ unchanged. This is true, whatever
the value for the halo size $L$. For the sake of completeness, we should note that this also
affects the source term: the source abundances are now $q^{\rm new}= x\times q^{\rm ref}$.

\section{Influence of the source spectrum\label{sec:src}}

\begin{table}[!t]
\caption{Best-fit transport parameters for different slopes $\gamma$.}
\label{tab:systematics_gamma}
\centering
\begin{tabular}{lcccccc} \hline\hline
$\gamma$  & $\!\!K_0^{\rm best}\times 10^2\!\!$     & $\delta^{\rm best}$ &  $V_c^{\rm best}$  &  $V_a^{\rm best}$  & $\!\!\!\!\chi^2$/d.o.f$\!\!\!\!$   \\
        & $\!\!\!\!\!\!$(kpc$^2\,$Myr$^{-1}$)$\!\!\!\!\!\!$ &                     &  $\!\!$(km$\;$s$^{-1}$)$\!\!$  & $\!\!$(km$\;$s$^{-1}$)$\!\!$   & \\\hline
& \multicolumn{1}{c}{} \vspace{-0.20cm} \\ 
0:~$2.55$         &    4.13    &    0.39    &    \dots   &    \dots   &    29.7    \vspace{0.05cm}\\  
{\it 0:}~${\it 2.65}$   & {\it 4.08} & {\it 0.40} & {\it \dots}& {\it \dots}& {\it 28.8} \vspace{0.05cm}\\  
0:~$2.75$         &    4.02    &    0.40    &    \dots   &    \dots   &    27.9    \vspace{0.05cm}\\  
0:~$2.85$         &    3.97    &    0.40    &    \dots   &    \dots   &    27.0    \vspace{0.25cm}\\  
I:~$2.55$         &    0.42    &    0.93    &    14.0    &    \dots   &    11.4    \vspace{0.05cm}\\  
{\it I:}~${\it 2.65}$   & {\it 0.42} & {\it 0.93} & {\it 13.5} & {\it \dots}& {\it 11.2} \vspace{0.05cm}\\  
I:~$2.75$         &    0.42    &    0.93    &    13.1    &    \dots   &    11.0    \vspace{0.05cm}\\  
I:~$2.85$         &    0.42    &    0.93    &    12.7    &    \dots   &    10.9    \vspace{0.25cm}\\  
II:~$2.55$        &    9.41    &    0.24    &    \dots   &    74.2    &    4.52    \vspace{0.05cm}\\  
{\it II:}~${\it 2.65}$  & {\it 9.76} & {\it 0.23} & {\it \dots}& {\it 73.1} & {\it 4.73} \vspace{0.05cm}\\  
II:~$2.75$        &    10.1    &    0.23    &    \dots   &    72.1    &    4.94    \vspace{0.05cm}\\  
II:~$2.85$        &    10.4    &    0.22    &    \dots   &    71.0    &    5.15    \vspace{0.25cm}\\  
III:~$2.55$       &    0.52    &    0.84    &    18.9    &    39.7    &    1.25    \vspace{0.05cm}\\  
{\it III:}~${\it 2.65}$ & {\it 0.48} & {\it 0.86} & {\it 18.8} & {\it 38.0} & {\it 1.47} \vspace{0.05cm}\\  
III:~$2.75$       &    0.45    &    0.87    &    18.8    &    36.4    &    1.75    \vspace{0.05cm}\\  
III:~$2.85$       &    0.42    &    0.88    &    18.6    &    35.0    &    2.10    \vspace{0.10cm}\\  
\hline
\end{tabular}
\note{\tiny Models 0, I, II, and III for $L=4$~kpc. The parameter $\gamma$ is the slope of the
propagated fluxes ($\gamma\!=\!\alpha+\delta$). The source spectrum is $Q(E)\propto \beta^{\eta_S}
{\cal R}^{-\alpha}$, and $\eta_S=-1$.\vspace{-0.25cm}}
\end{table}

In Paper~I, we showed that some correlations exist between the source spectra and the
transport parameters. Below, we investigate in more detail how the transport
parameters are changed if we assume different input for $\alpha$ (spectral index of the sources)
and $\eta_S$ (related to the spectral shape at low energy:
\begin{equation}
 Q(E)\propto \beta^{\eta_S} {\cal R}^{-\alpha}.
 \label{eq:src}
\end{equation}

	\subsection{Influence of the source slope $\alpha$}

It is usually said that the source spectrum energy dependence $Q(E)$ factors out of the
secondary-to-primary ratio. This agrees with the results gathered in
Table~\ref{tab:systematics_gamma}, which shows the effect of varying the source slope $\alpha$
(or equivalently the slope of the measured fluxes $\gamma=\alpha+\delta$). This is especially
true for Model~I (no reacceleration), whereas if reacceleration is included (Models~II and~III),
a few \% change on the parameter $\gamma$ leads to a few \% change on  $V_a$, $K_0$, and
$\delta$. 

From the last column of the same table, we see that on a statistical basis, the best model
is Model~III, then Model~II, Model~I, and finally Model~0. The relative merit (goodness-of-fit)
of each class of models (III, II, I, and 0) remains unchanged when $\alpha$ is changed.

	\subsection{Influence of the low-energy source spectrum through $\eta_S$}
\begin{figure}[!t]
\centering
\includegraphics[width = \columnwidth]{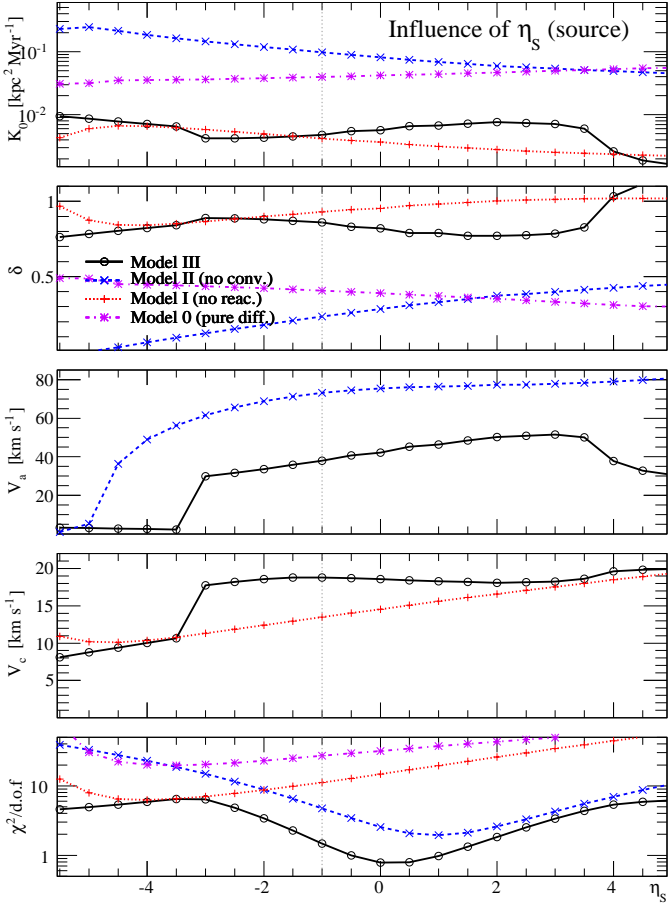}
\caption{From top to bottom: best-fit values of $K_0$, $\delta$, $V_a$, $V_c$, and the
associated $\chi^2/d.o.f.$ as a function of $\eta_S$ [see Eq.~(\ref{eq:src})].
Here, $\gamma=\alpha+\delta=2.65$. The vertical grey-dotted line is a guideline
for the results at the default configuration $\eta_S=-1$.}
\label{fig:param_vs_EtaS}
\end{figure}
The second parameter that can be varied in the source term is the low-energy shape
$\beta^{\eta_S}$ [see Eq.~(\ref{eq:src})], which is {\em a
priori} not very well constrained, both from observational and theoretical points of view.
The evolution of the best-fit transport parameters as a function of $\eta_S$ is shown in
Fig.~\ref{fig:param_vs_EtaS}, for Model 0 (dash-dotted lines/stars), I
(dotted lines/plusses), II (dashed lines/crosses), and III (solid lines/open circles).

The three configurations have marked minima in their $\chi^2_{\rm min}$ (bottom panel), but
not at the same $\eta_S$. Models~0 and~I have a similar $\chi^2$ dependence (weak impact
of the presence of the wind), as have Models~II and~III (both affected by reacceleration).
Understanding the exact dependence of the transport parameters with $\eta_S$ is not the goal of
the analysis. The important point to
underline is that changing $\eta_S$ does not allow reconciling convection Models~I and~III
with smaller $\delta$. As $\delta$ varies only in the range $0.3-0.5$ for Model~0 (pure diffusion), we
conclude that for present B/C data, large $\delta$ are associated to the constant Galactic wind.
The largest variation is observed for Model~II (reacceleration only), for which $\delta$ is allowed to be
either smaller or larger than $\delta^{\rm best}_{\rm II}=0.33$ (associated with $\eta^{\rm
best}_S\approx 1)$ for a smaller or a larger $\eta_S$. For any $\eta_S$,
Model~III always performs better than the other two models.

	\subsection{Fitting $\alpha$ and $\eta_S$ using a primary flux}

The source parameters $\alpha$ and $\eta_S$ are not completely free.  They can be fitted
directly on primary elemental fluxes (see Paper~I). 
If we determine simultaneously the transport and source parameters from
a fit on B/C + O HEAO-3 data, we obtain the best-fit values gathered in Table~\ref{tab:best_fit_B/C+O}.
For Models~II and~III, $\eta_S^{B/C+O}$ is not far from $\eta_S^{\rm best}$ found from the B/C
fit (see Fig.~\ref{fig:param_vs_EtaS}). For Model~II, using the above constraint 
($\eta_S^{B/C+O}=1.24$) selects the diffusion slope $\delta^{B/C+O}=0.35$, close to the
Kolmogorov spectrum for the turbulence. 
\begin{table}[!t]
\caption{Best-fit values from a simultaneous fit on B/C and O.}
\label{tab:best_fit_B/C+O}
\centering
\begin{tabular}{lcccc} \hline\hline
Model & $\eta_S^{(B/C+O)}$  & $\alpha^{(B/C+O)}$ & $\delta^{(B/C+O)}$ & $\chi^2_{\rm min}/$d.o.f.\vspace{0.10cm}   \\\hline
& \multicolumn{1}{c}{} \vspace{-0.20cm} \\
0     &      $-0.82$        &        2.22        &        0.41        & 25.3  \vspace{0.cm} \\
I     &      $-1.23$        &        2.23        &        0.98        & 9.23  \vspace{0.cm} \\
II    &      $1.24$         &        2.24        &        0.35        & 3.62  \vspace{0.cm} \\ 
III   &      $1.56$         &        2.30        &        0.95        & 2.31  \vspace{0.10cm} \\
\hline
\end{tabular}
\note{\tiny The data used are IMP7-8, Voyager~1\&2, ACE, HEA0-3, Spacelab, and CREAM04 for the B/C ratio, and HEAO-3 for the oxygen flux.\vspace{-0.25cm}}
\end{table}
%

The surprising result is that the value for $\alpha^{B/C+O}$ is quite
resilient to any slope obtained for $\delta$. This is odd, since it leaves the propagated slope
$\gamma=\alpha+\delta$ quite unconstrained, leading to either 2.6 (Model~II) or 3.2 (Model~III).
This indicates that in Model~III, pure diffusive transport (i.e. $\gamma=\alpha+\delta$)
has not yet been reached at TeV/n energies.

	\subsection{Summary for the source effect}

The dependence of the transport parameters with the source parameters $\alpha$
and $\eta_S$ are very different. 
We checked that the transport parameters do not significantly depend on the
source parameter $\alpha$ (in particular, $\delta$ is unaffected). Still, correlations
exist between the source and the other transport parameters (see Paper~I), and from
Table~\ref{tab:systematics_gamma}, we can estimate that they affect their best-fit values at
most by a few ten percent.

On the other hand, Fig.~\ref{fig:param_vs_EtaS} shows that the transport parameters
(including $\delta$) are quite sensitive to the low-energy source spectrum parameterised by
$\eta_S$. In any case, this does not allow reconciliation of Model~III with small $\delta$. Taking
a sizeable range for $\eta_S$ always lead to $\delta\gtrsim 0.6$. This is even worse for
Model~I (no reacceleration). At variance, Model~II (no convection) is very sensitive to
$\eta_S$. However, $\eta_S$ is constrained by primary fluxes measurements, and by no means
could it span the range shown in Fig.~\ref{fig:param_vs_EtaS}. When this constraint is considered,
the best-fit $\delta$ for Model~II turns out to be $0.35$. Given the sensitivity of the latter
to $\eta_S$ and the possible bias in low-energy data (because of the solar modulation), this
is consistent with a Kolmogorov spectrum. If the constraint of the primary flux is taken into
account, the uncertainty on the transport parameters (associated to the source parameter
$\eta_S$) can be estimated as $\sim 10-20\%$.

The puzzling point is that, although Model~III is preferred over Model~II on a statistical
basis, the latter seems to agree more with the expected theoretical value of $\delta$.
This is more of further if we consider the quantity $\gamma=\alpha+\delta$, the asymptotic slope reached
in the purely diffusive regime. The TRACER experiment finds $\gamma\approx 2.65$ for all
nuclei~\citep{2008ApJ...678..262A}, whereas the best-fit slope from high-energy H and He 
data is $\gamma \approx 2.85$ \citep{2009PhRvL.102g1301D}. From the propagated slope's point of
view, Model~II is again favoured over Model~III, since the former leads to  $\gamma=2.59$,
whereas the latter leads to $3.25$ (read off Table~\ref{tab:best_fit_B/C+O}). If Model~III is
confirmed, as already emphasised above, this would mean that even high-energy data have not reached
the purely diffusive transport yet.

\section{Influence of the low-energy diffusion coefficient\label{sec:diff}}

We now turn to the effect of the low-energy shape of the diffusion coefficient. Assuming
different diffusion schemes lead to different forms of the spatial and momentum diffusion
coefficient~\citep{2002cra..book.....S}. Recently, \citet{2006ApJ...642..902P} also argue
that the form of the spatial diffusion coefficient can change at low energy, due to the
possibility that the nonlinear MHD cascade sets the power-law spectrum of turbulence. 

We first consider several theoretically-motivated diffusion forms given in the literature.
They mostly differ at low energy, so that it is useful to rewrite Eq.~(\ref{eq:K(E)}) as
\begin{equation}
K(E)= \beta^{\eta_T} \cdot K_0 {\cal R}^\delta \;.
\label{eq:eta_T}
\end{equation}
In the second step, we let $\eta_T$ vary over a wide range in order to draw more general
conclusions.

	\subsection{Influence of the turbulence scheme}

\begin{table}[!t]
\caption{$K(E)$ and $K_{\rm pp}$ for different schemes.}
\label{tab:Kxx_diff_cases}
\centering
\begin{tabular}{llcc} \hline\hline
\multicolumn{2}{l}{Type of turbulence}&    $\eta_T$    &   $\frac{K_{pp} K_{\rm xx}}{4/3\, p^2V_a^2}$ \vspace{0.10cm}   \\\hline
& \multicolumn{1}{c}{} \vspace{-0.20cm} \\ 
LBI&Leaky Box Inspired                &       0        &  $\frac{1}{\delta\,(4-\delta^2)\,(4-\delta)}$  \vspace{0.15cm} \\
{\it SA}&{\it Slab Alfv\'en}          &     {\it 1}    &  $\frac{1}{\delta\,(4-\delta^2)\,(4-\delta)}$  \vspace{0.15cm} \\ 
IFM&Isotropic fast magnetosonic       &  $\!\!\!\!{2\!-\!\delta}\!\!\!\!$  &  $\beta^{1-\delta} \ln (\frac{v}{V_a})\!\!\!$   \vspace{0.15cm} \\
Mix& Mixture SA and IFM               &  $\!\!\!\!{1\!-\!\delta}\!\!\!\!$  &  $\beta^{1-\delta} \ln (\frac{v}{V_a})$  \vspace{0.1cm} \\
\hline
\end{tabular}
\note{\tiny The spatial diffusion coefficient is $K_{\rm xx}= \beta^{\eta_T} \cdot K_0 \cdot {\cal R}^\delta$.\vspace{-0.25cm}}
\end{table}
A few turbulence schemes are gathered in Table~\ref{tab:Kxx_diff_cases}. The associated best-fit values
of the transport coefficients are presented in Table~\ref{tab:systematics_coefdif}.
\begin{table}[!t]
\caption{Best-fit transport parameters based on different low-energy dependence of the
diffusion coefficient.}
\label{tab:systematics_coefdif}
\centering
\begin{tabular}{lcccccc} \hline\hline
Type    & $\!\!\!K_0^{\rm best}\times 10^2\!\!$     & $\delta^{\rm best}$ &  $V_c^{\rm best}$  &  $V_a^{\rm best}$  & $\!\!\!\!\chi^2$/d.o.f$\!\!\!\!$   \\
        & $\!\!\!\!\!\!\!\!\!\!\!$(kpc$^2\,$Myr$^{-1}$)$\!\!\!\!\!\!\!\!\!\!$ &                     &  $\!$(km$\;$s$^{-1}$)$\!$  & $\!$(km$\;$s$^{-1}$)$\!$   & \\\hline
& \multicolumn{1}{c}{} \vspace{-0.20cm} \\ 
0:~LBI       &   3.48   &   0.45   &  \dots   &  \dots   &   17.5     \vspace{0.05cm}\\  
{\it 0:~SA  }&{\it 4.08}&{\it 0.40}&{\it\dots}&  \dots   &{\it 28.8}  \vspace{0.05cm}\\  
0:~IFM       &   4.30   &   0.38   &  \dots   &  \dots   &   36.7     \vspace{0.05cm}\\  
0:~Mix       &   3.71   &   0.43   &  \dots   &  \dots   &   23.7    \vspace{0.25cm}\\  
%
I:~LBI       &   0.40   &   0.94   &   13.6   &  \dots   &   12.0     \vspace{0.05cm}\\  
{\it I:~SA  }&{\it 0.42}&{\it 0.93}&{\it 13.5}&{\it\dots}&{\it 11.2}  \vspace{0.05cm}\\  
I:~IFM       &   0.42   &   0.93   &   13.5   &  \dots   &   11.6     \vspace{0.05cm}\\  
I:~Mix       &   0.41   &   0.94   &   13.5   &  \dots   &   12.0    \vspace{0.25cm}\\  
II:~LBI      &   5.50   &   0.38   &  \dots   &   65.0   &   1.61     \vspace{0.05cm}\\  
{\it II:~SA }&{\it 9.76}&{\it 0.23}&  \dots   &{\it 73.1}&{\it 4.73}  \vspace{0.05cm}\\  
II:~IFM      &   14.0   &   0.16   &  \dots   &   18.9   &   6.86     \vspace{0.05cm}\\  
II:~Mix      &   7.13   &   0.32   &  \dots   &   12.8   &   2.03     \vspace{0.25cm}\\  
III:~LBI     &   0.70   &   0.78   &   18.0   &   47.1   &   0.87     \vspace{0.05cm}\\  
{\it III:~SA}&{\it 0.48}&{\it 0.86}&{\it 18.8}&{\it 38.0}&{\it 1.47}  \vspace{0.05cm}\\  
III:~IFM     &   0.49   &   0.85   &   18.9   &   45.6   &   1.25     \vspace{0.05cm}\\  
III:~Mix     &   0.73   &   0.77   &   17.8   &   57.4   &   0.93     \vspace{0.10cm}\\  
\hline
\end{tabular}
\note{\tiny Models 0, I, II, and III for $L=4$~kpc. SA corresponds to the reference
DM used throughout the paper.\vspace{-0.25cm}}
\end{table}
When only convection is present (Model~I), the low-energy form of $K(E)$ is irrelevant, as
seen in Table~\ref{tab:systematics_coefdif}. For models with
reacceleration (Models~II and~III), it significantly affects almost all parameters. If there is
no wind (Model~II), the effect is maximum on $\delta$. The model with pure diffusion (Model~0),
or with both convection and reacceleration (Model~III), falls in-between. For Model III, the
cases SA and IFM, on the one hand, and LBI and Mix, on the other, give very similar results.
This is easily understood as the quantities $\eta_T^{\rm SA}=2-\delta$ and $\eta_T^{\rm IMF}=1$
(respectively $\eta_T^{\rm LBI}=1-\delta$ and $\eta_T^{\rm Mix}=0$) are roughly equal
for $\delta_{\rm III}^{\rm best}\sim 1$.

The important result is that Model~III is mildly sensitive to the diffusion scheme,
with an uncertainty of a few ten percent scatter on all the transport parameters.
However, Model~II parameters are extremely sensitive to the diffusion scheme (more than
a factor of 2 scatter). Depending on the case considered, $\delta$ is found in the range
$0.16-0.38$. The hierarchy of $\chi^2_{\rm min}$ among the various models is always conserved.

	\subsection{Generalisation to any $\eta_T$}

We generalise the analysis by allowing for any value of the parameter $\eta_T$ in the diffusion
coefficient $K(E)$. In doing so, we do not seek to provide sound physical motivations for
the range tested. In this section, whatever the value of $\eta_T$, the diffusion
coefficient in momentum space is assumed to follow
$K_{pp} K_{\rm xx} = (4/3)\, p^2V_a^2 /(\delta\,(4-\delta^2)\,(4-\delta))$. 

\begin{figure}[!t]
\centering
\includegraphics[width = \columnwidth]{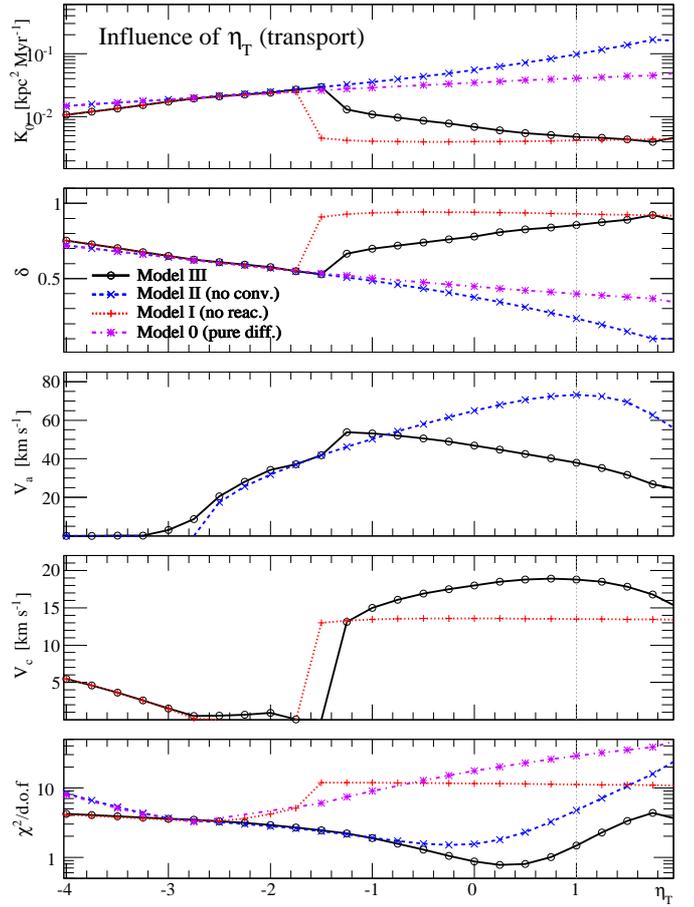}
\caption{Same as Fig.~\ref{fig:param_vs_EtaS}, but now as a function of $\eta_T$
[see Eq.~(\ref{eq:eta_T})]. The vertical grey-dotted line is for the default
configuration $\eta_T=1$ (SA).}
\label{fig:param_vs_EtaT}
\end{figure}
The best-fit transport parameters and $\chi^2_{\rm min}$ evolution as a function of $\eta_T$ are
plotted in Fig.~\ref{fig:param_vs_EtaT}. For $\eta_T\lesssim -2$, all the models converge slowly
towards purely diffusive models (no convection, no reacceleration). But this is at the cost of a
bad $\chi^2$ (see bottom panel). Based on the $\chi^2$ criterium, high values of $\eta_T$ ($\gtrsim 2$)
are also disfavoured. The four configurations have marked minima in their $\chi^2_{\rm min}$,
corresponding to $\eta^{\rm best}_T\approx -2.75,-2.5,-0.25,+0.25$ for Models 0, I, II, and III
respectively, for which $\delta^{\rm best}\approx 0.6,\,0.6,\,0.4,\,0.8$. For $\eta_T\sim -2.5$,
all models point to $\delta\sim 0.5$, close to a Kraichnan spectrum for turbulence.

For Model~III, for a well-chosen value of $\eta_T$, the diffusion slope could be
decreased at most down to $\delta=0.5$, but such a configuration does not correspond to the
minimal $\chi^2$ for this model so is excluded. For Model~II, almost any value of $\delta$
can be reached.

	\subsection{Summary for the diffusion coefficient effect}

Similars to the effect of the source parameter $\eta_S$, varying the parameter $\eta_T$ on a
wide range i) does not to allow $\delta\lesssim 1/2$ to reached for Model~III, and ii) strongly
affects $\delta$ for Model~II. For the latter, the best-fit $\eta_T$ leads to $\delta\approx0.4$,
slightly more than the Kolmogorov spectrum of turbulence.

From a statistical point of view, Model~III is again preferred, except in the region $-3\lesssim
\eta_T\lesssim -1.5$ where $V_c$ drops to zero, so that it is equivalent to Model~II. For 
$\eta_T\lesssim -2.75$, $V_a$ also drops to zero, so that all models are close to the purely
diffusive case (Model~0), which favours $\delta\sim 1/2$. As before, one of the interesting
features of Model~II (reacceleration only) is its versatility. As the range of allowed $\eta_T$
remains unspecified to some extent, Model~II does not point to any specific value of $\delta$.
It can accommodate values as low as $1/3$, going through $1/2$ and even higher values, if a
sharp turn-off (negative values of $\eta_T$) exists in the diffusion coefficient (e.g.,
\citealt{2006ApJ...642..902P}).

\section{Influence of the production cross-sections\label{sec:Xsec}}

Many reaction channels are required to calculate the production of secondary species.
Semi-analytical formulae, semi-empirical approaches, or even fit to the data are
currently used in the literature to obtain the full set of production cross-section in CR
physics.
We show first the different best-fit values obtained for different sets available in the literature.
We then inspect what would be the effect of energy-biased cross-sections on the transport
parameter derivation.

\subsection{Using different sets of fragmentation cross-sections}

We use four different sets of cross-sections. Among them, the most up-to-date are W03
\citep{2003ApJS..144..153W} and GAL09 (from the GALPROP code\footnote{\tiny http://galprop.stanford.edu/web\_galprop/galprop\_home.html}). The former is  based on the
semi-empirical approach of Webber and coworkers, initiated in the 90's
\citep{1990PhRvC..41..566W}. The latter set takes advantage of the former approach,
renormalising some cross-sections to selected data. The two others, WKS98
\citep{1998ApJ...508..940W} and S01 (private communication of Aim\'e Soutoul, 2001), are a
bit outdated, but they serve as an illustration here. 

The effect of these different sets on the parameter determination is shown in
Table~\ref{tab:systematics_xsec1}.
\begin{table}[!t]
\caption{Best-fit transport parameters for various cross-section sets.}
\label{tab:systematics_xsec1}
\centering
\begin{tabular}{lcccccc} \hline\hline
X-files$\!\!$  & $K_0^{\rm best}\times 10^2$     & $\delta^{\rm best}$ &  $V_c^{\rm best}$  &  $V_a^{\rm best}$  & $\!\!\!\!\chi^2$/d.o.f$\!\!\!\!$   \\
               & $\!\!\!\!\!\!\!$(kpc$^2\,$Myr$^{-1}$)$\!\!\!\!\!\!\!$ &                     & $\!\!$ (km$\;$s$^{-1}$)  & $\!\!\!$(km$\;$s$^{-1}$)$\!\!\!$   & \\\hline
& \multicolumn{5}{c}{} \vspace{-0.25cm}\\
0:~WKS98$\!\!\!\!\!\!\!\!\!\!$    &   2.66   &   0.53   &   \dots   &   \dots   &    23.5    \vspace{0.05cm}\\  
0:~S01$\!\!\!$                    &   3.40   &   0.40   &   \dots   &   \dots   &    34.5    \vspace{0.05cm}\\  
{\it 0:~W03}$\!\!\!\!\!$          &{\it 4.08}&{\it 0.40}&{\it \dots}&{\it \dots}& {\it 28.8} \vspace{0.05cm}\\  
0:~GAL09$\!\!\!\!\!\!\!\!\!\!$    &   3.83   &   0.46   &   \dots   &   \dots   &    28.4    \vspace{0.25cm}\\  
I:~WKS98$\!\!\!\!\!\!\!\!\!\!$    &   0.45   &   0.95   &   10.4    &   \dots   &    12.0    \vspace{0.05cm}\\  
I:~S01$\!\!\!$                    &   0.25   &   1.01   &   11.9    &   \dots   &    13.5    \vspace{0.05cm}\\  
{\it I:~W03}$\!\!\!\!\!$          &{\it 0.42}&{\it 0.93}&{\it 13.5} &{\it \dots}& {\it  11.2} \vspace{0.05cm}\\  
I:~GAL09$\!\!\!\!\!\!\!\!\!\!$    &   0.49   &   0.95   &   13.6    &   \dots   &    12.3    \vspace{0.25cm}\\  
II:~WKS98$\!\!\!\!\!\!\!\!\!\!$   &   7.19   &   0.31   &   \dots   &   71.5    &    3.40    \vspace{0.05cm}\\  
II:~S01$\!\!\!$                   &   9.27   &   0.22   &   \dots   &   68.8    &    6.56    \vspace{0.05cm}\\  
{\it II:~W03}$\!\!\!\!\!$         &{\it 9.76}&{\it 0.23}&{\it \dots}&{\it 73.1} & {\it  4.73} \vspace{0.05cm}\\  
II:~GAL09$\!\!\!\!\!\!\!\!\!\!$   &   10.0   &   0.26   &   \dots   &   85.0    &    4.03    \vspace{0.25cm}\\  
III:~WKS98$\!\!\!\!\!\!\!\!\!\!$  &   0.69   &   0.80   &   15.8    &   43.5    &    1.11    \vspace{0.05cm}\\  
III:~S01$\!\!\!$                  &   0.25   &   0.98   &   16.2    &   27.9    &    3.01    \vspace{0.05cm}\\  
{\it III:~W03}$\!\!\!\!\!$        &{\it 0.48}&{\it 0.86}&{\it 18.8} &{\it 38.0} &{ \em 1.47} \vspace{0.05cm}\\  
III:~GAL09$\!\!\!\!\!\!\!\!\!\!$  &   0.65   &   0.82   &   21.7    &   49.4    &    1.53    \vspace{0.10cm}\\  
\hline
\end{tabular}
\note{\tiny Models 0, I, II, and III for $L=4$~kpc. 
The four base sets of cross-sections are WKS98 from \citet{1998ApJ...508..940W},
S01 from a private communication of Aim\'e Soutoul (2001), W03 from \citet{2003ApJS..144..153W},
and GAL09 from the GALPROP code v50.1p (2009).\vspace{-0.25cm}}
\end{table}
The scatter in the values of the different parameters is not the same depending on the model
considered. WKS98 requires less convection than any other set, and GAL09 needs more
reacceleration than the others. Otherwise, it is difficult to find clear trends. For
instance, the W03 and GAL09 sets give very similar results for Model~I, but not as similar as
for Models~II and~III. 

If we discard the S01 set (which is based on an unpublished preliminary analysis),
we can conclude that the {\em SystUnc} on the transport parameters related to the choice
of the production cross-section is $\lesssim 20\%$. This figure is similar to the
uncertainty quoted for the  parameterisation of the cross-section themselves ($\sim
10-20\%$).

\subsection{Influence of a systematic energy bias in W03 cross-sections}

\begin{figure}[!t]
\centering
\includegraphics[width = \columnwidth]{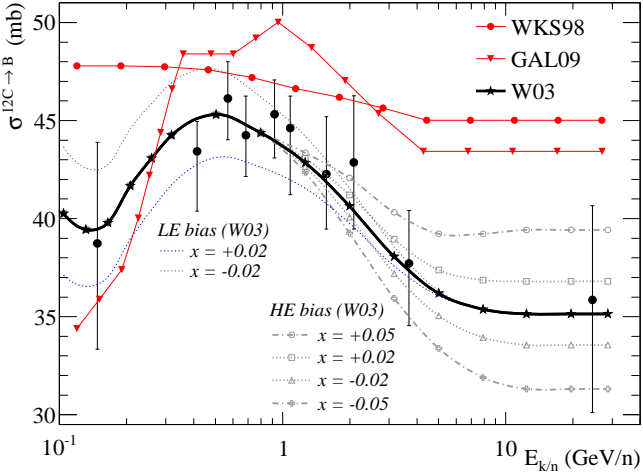}
\caption{Production cross-section for $^{12}$C+H$\rightarrow ^{10,11}$B (adapted from
\citealt{2003ApJS..144..153W}). The standard sets are shown as solid lines (WKS98: red dots;
GAL09: red down triangles; W03: black stars), and the biased sets in dotted ($|x|=0.02$)
and dashed ($|x|=0.05$) lines.}
\label{fig:Xsec}
\end{figure}

A striking result of the updated cross-section formulation of W03 \citep{2003ApJS..144..153W}
compared to WKS98 \citep{1998ApJ...508..940W} was a systematic energy bias (smaller
cross-sections at higher energy). 
To address the effect of a possible {\em residual} bias, we allow for a systematic
energy bias in all production cross-sections, either at low energy (LE)
or at high energy (HE). Biased sets are obtained from the reference set W03 by 
applying a factor $(E_{\rm k/n}/10)^{-x}$ below 10 GeV/n (LE bias), and $(E_{\rm k/n})^{x}$
above 1~GeV/n (HE bias, but left constant above 10 GeV/n).
This is illustrated for the $^{12}$C$\rightarrow (^{10}$B$+^{11}$B)
cross-section, as shown in Fig.~\ref{fig:Xsec}. As seen from this figure and the
experimental data, biases $x$ as large as $0.05$ are marginally consistent with the data
at low energy, but are acceptable at high energy.

In Table~\ref{tab:Systematics_xsecHE}, we only report the results of the high-energy
biases on the best-fit values of the transport parameters. We checked that the result of a
low-energy or a high-energy bias gives the same trends.
\begin{table}[!t]
\caption{High-energy biases on the W03 cross-section set.}
\label{tab:Systematics_xsecHE}
\centering
\begin{tabular}{lccccc} \hline\hline
$\!$W03$\!$  & $\!\!K_0^{\rm best}\times 10^2\!\!$     & $\delta^{\rm best}$ &  $\!V_c^{\rm best}\!$  &  $\!\!V_a^{\rm best}\!\!$  & $\!\!\!\!\chi^2$/d.o.f$\!\!\!\!$   \\
$\!\!\!$HE bias $x$    & $\!\!\!\!\!\!\!\!$(kpc$^2\,$Myr$^{-1}$)$\!\!\!\!\!\!$ &                     &  $\!$(km$\;$s$^{-1}$)$\!\!$   & $\!\!\!$(km$\;$s$^{-1}$)$\!\!$   & \\\hline
& \multicolumn{5}{c}{} \vspace{-0.25cm}\\
0:~$+0.05$          &   4.05   &   0.45   &   \dots   &   \dots   &   26.5    \vspace{0.05cm}\\  
0:~$+0.02$          &   4.06   &   0.42   &   \dots   &   \dots   &   27.7    \vspace{0.05cm}\\  
{\it 0:~}${\it +0.00}$&{\it 4.08}&{\it 0.40}&{\it \dots}&{\it \dots}& {\it 28.8} \vspace{0.05cm}\\  
0:~$-0.02$          &   4.09   &   0.38   &   \dots   &   \dots   &   30.1    \vspace{0.05cm}\\  
0:~$-0.05$          &   4.13   &   0.34   &   \dots   &   \dots   &   32.5    \vspace{0.25cm}\\  
I:~$+0.05$          &   0.61   &   0.89   &   13.7    &   \dots   &   12.1    \vspace{0.05cm}\\  
I:~$+0.02$          &   0.49   &   0.91   &   13.6    &   \dots   &   11.4    \vspace{0.05cm}\\  
{\it I:~}${\it +0.00}$&{\it 0.42}&{\it 0.93}&{\it 13.5} &{\it \dots}&{\it 11.2} \vspace{0.05cm}\\  
I:~$-0.02$          &   0.35   &   0.95   &   13.4    &   \dots   &   11.2    \vspace{0.05cm}\\  
I:~$-0.05$          &   0.27   &   0.98   &   13.1    &   \dots   &   11.8    \vspace{0.25cm}\\  
II:~$+0.05$         &  9.20    &   0.28   &   \dots   &   78.0    &   3.26    \vspace{0.05cm}\\  
II:~$+0.02$         &  9.51    &   0.25   &   \dots   &   75.0    &   4.13    \vspace{0.05cm}\\  
{\it II:~}${\it +0.00}$&{\it 9.76}&{\it 0.23}&{\it \dots}&{\it 73.1} &{\it 4.73} \vspace{0.05cm}\\  
II:~$-0.02$         &  10.0    &   0.21   &   \dots   &   71.2    &   5.35    \vspace{0.05cm}\\  
II:~$-0.05$         &  10.4    &   0.19   &   \dots   &   68.1    &   6.27    \vspace{0.25cm}\\  
III:~$+0.05$        &  0.98    &   0.74   &   19.0    &   50.0    &   1.01    \vspace{0.05cm}\\  
III:~$+0.02$        &  0.75    &   0.81   &   19.0    &   42.7    &   1.22    \vspace{0.05cm}\\  
{\it III:~}${\it +0.00}$ &{\it 0.48}&{\it 0.86}&{\it 18.8} &{\it 38.0} &{\it 1.47} \vspace{0.05cm}\\  
III:~$-0.02$        &  0.35    &   0.91   &   18.5    &   33.4    &   1.88    \vspace{0.05cm}\\  
III:~$-0.05$        &  0.21    &   1.01   &   17.8    &   26.6    &   2.94    \vspace{0.10cm}\\  
\hline
\end{tabular}
\note{\tiny Models 0, I, II, and III for $L=4$~kpc.
The various sets are obtained after multiplying the W03 set by $(E_{\rm k/n})^{x}$
above 1~GeV/n (and kept constant above 10 GeV/n), with $x=+0.05,\,+0.02,\,0,\,-0.02,\,-0.05$.
(corresponding to bottom to top curves on the right-hand side of Fig.~\ref{fig:Xsec}).\vspace{-0.25cm}}
\end{table}
Quite naturally, there is a significant correlation of the bias with the transport parameters.
We note that if $\delta$ increases, all the other transport parameters $K_0$, $V_a$, and $V_c$
decrease. In principle, we would expect that $\delta$ increases if the bias is positive: larger
cross-sections at high-energy produce more secondaries at high-energy, requiring a larger
$\delta$ to match the same B/C data. This is observed for Models~0 and~II. However, the reverse
effect is observed for Models~I and III, i.e., for models with convective wind.
\begin{figure*}[!t]
\centering
\includegraphics[width = \columnwidth]{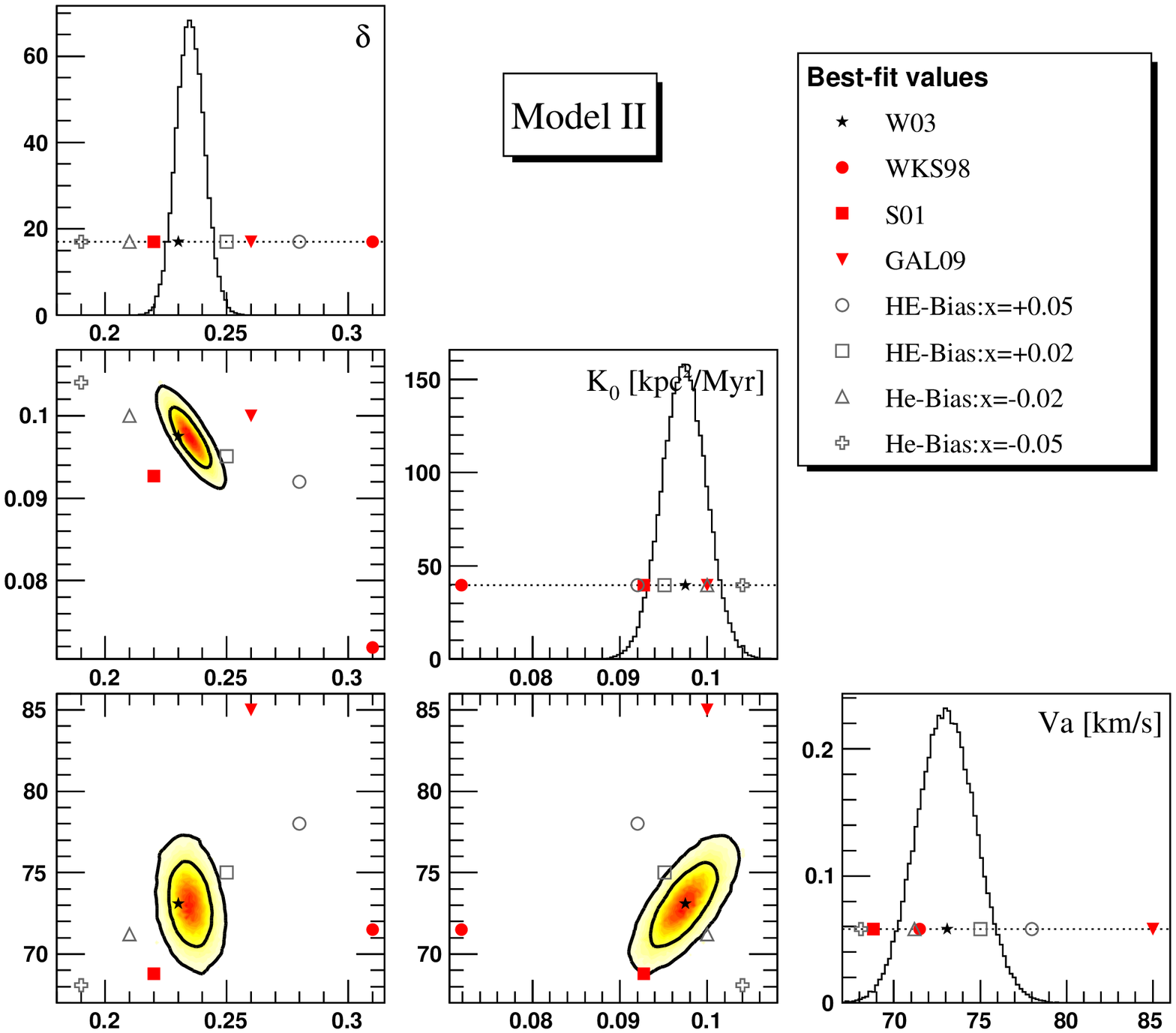}
\includegraphics[width = \columnwidth]{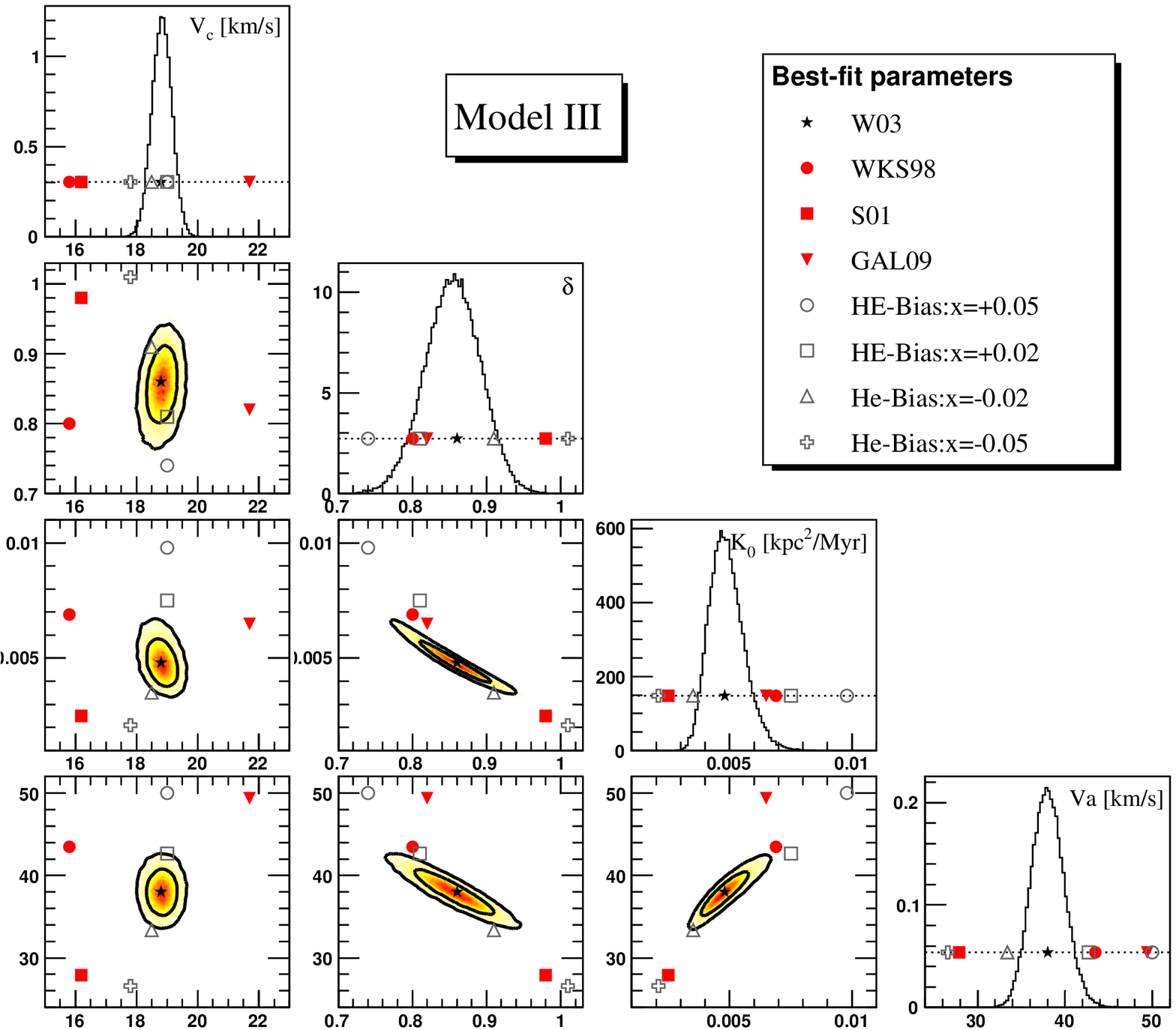}
\caption{Left panel: Model~II (pure reacceleration). Right panel: Model~III (reacceleration
and convection). PDF of the transport parameters (for the reference inputs) as obtained in
Paper~II, along with the best-fit values for different production cross-section sets.}
\label{fig:pdfs}
\end{figure*}

Such an exercise has limitations since it is not realistic to expect such systematic
energy biases for all production channels. However, it gives some trend and further
confirms that Model~III cannot be reconcile with small $\delta$.

\subsection{Summary for the influence of fragmentation cross-sections}

The results when using different production cross-section sets are difficult to interpret.
To check that the observed trends do not come from an error in the code, we compared
our results with the LBM analysis of \citet{2003ApJS..144..153W}. These authors find a
{\em significantly smaller escape length dependence on rigidity (from $P^{0.6}$ to
$P^{0.5}$)}, going from the WKS98 to the W03 set. From both an LBM analysis with our
propagation code and a DM analysis using the LBI case (see Table~\ref{tab:Kxx_diff_cases})
with the addition of a rigidity cut-off (not discussed in this paper), we also find this
decrease of $\Delta\delta=0.1$. This is also confirmed for the more standard Model~0
(pure diffusion) where we go from 0.53 to 0.40.

We thus have to conclude, as seen from Table~\ref{tab:systematics_xsec1}, that the effect of
these cross-section sets is not the same for different classes of models: it strongly depends  on
the presence or absence of convection and, to a lesser extent, on reacceleration. Nevertheless,
for all models, the typical scatter in the best-fit values is a factor of 2 for $K_0$, $\sim
50\%$ for $V_a$, $\sim 10\%$ for $\delta$, and $\sim 5\%$ for $V_c$. 
Figure~\ref{fig:pdfs} illustrates that uncertainties in some input
parameters (here the production cross-sections) provide larger {\em SystUnc} on the transport
parameters than their {\em StatUnc} calculated using the MCMC technique (taken from Paper~II). 

As for the other effects inspected in previous sections, Model~III always provide $\delta$
larger than $\sim 0.6$. Model~II provides values of $\delta$ in the correct range for a
Kolmogorov spectrum, but it never does better than Model~III in terms of the goodness-of-fit.

\section{Influence of an energy bias in HEAO-3 B/C data\label{sec:biasHEAO}}

It is obvious that most of the results derived in this paper rely on the confidence we 
place in the HEAO-3 data, since they are the most constraining data to date. Their very small
error bars at high energy could over-constrain $\delta$. Several other studies
use large error bars for high energy B/C HEAO-3 points, contradicting the prescription
given in the original paper \citep{1990A&A...233...96E}, and this could bias their results.
As an exercise, we allowed for an energy bias in these data. This approach is disputable
so we do not wish to defend it strongly. We inspected whether a small bias in
the data strongly affects the best-fit values of the transport parameters.

\begin{figure}[!t]
\centering
\includegraphics[width = \columnwidth]{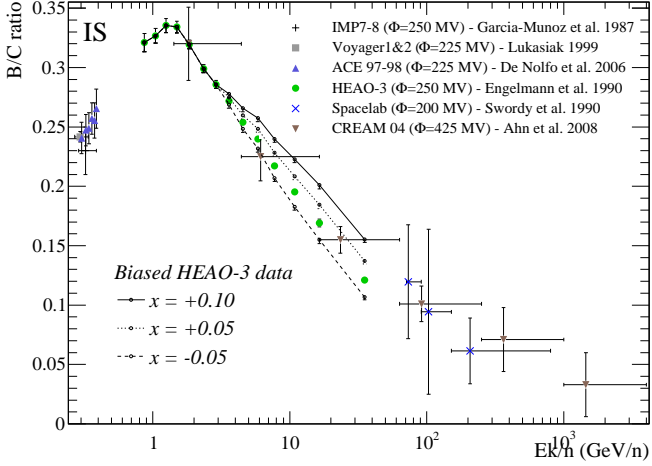}
\caption{Demodulated B/C data used in this paper (denoted dataset F).
Biased HEAO-3 data are shown as solid, dashed, and dotted lines.}
\label{fig:bias_data}
\end{figure}

The HEAO-3 data are biased above $E_c$, using $(E_{\rm k/n}/E_c)^{x}$ for $E_{\rm k/n}>E_c$,
with $E_c=$~2.9 GeV/n, as shown in Fig.~\ref{fig:bias_data}. 
As in previous sections, we fit Models~0, I, II, and III on dataset F (as shown in
Fig.~\ref{fig:bias_data}), where the HEAO-3 data have been replaced by the biased ones. 
The results have been gathered in Table~\ref{tab:Systematics_HEAO3}. Intuitively, in a similar
fashion as for the bias in the cross-sections, we expect that a higher value of the
B/C data at high energy (positive bias) leads to a smaller $\delta$. This is what
is observed for all models. The effect of the bias on $\delta$ is at its maximum for the pure
diffusion model (Model~0), where $\Delta\delta\sim 0.2$. For Models~II and III,
the effect is less pronounced owing to the presence of reacceleration, with a change
of only $\Delta\delta\sim 0.1$.
\begin{table}[!t]
\caption{Influence of biasing HEAO-3 data.}
\label{tab:Systematics_HEAO3}
\centering
\begin{tabular}{lccccc} \hline\hline
$\!$HEAO-3$\!$  & $\!\!K_0^{\rm best}\times 10^2\!\!$     & $\delta^{\rm best}$ &  $\!V_c^{\rm best}\!$  &  $\!\!V_a^{\rm best}\!\!$  & $\!\!\!\!\chi^2$/d.o.f$\!\!\!\!$   \\
$\!\!\!$HE bias $x$ & $\!\!\!\!\!\!\!\!$(kpc$^2\,$Myr$^{-1}$)$\!\!\!\!\!\!$ &                     &  $\!$(km$\;$s$^{-1}$)$\!\!$   & $\!\!\!$(km$\;$s$^{-1}$)$\!\!$   & \\\hline
& \multicolumn{5}{c}{} \vspace{-0.25cm}\\
0:~$+0.10$          &   4.97   &   0.28   &   \dots   &   \dots   &   22.5    \vspace{0.05cm}\\  
0:~$+0.05$          &   4.50   &   0.34   &   \dots   &   \dots   &   25.2    \vspace{0.05cm}\\  
{\it 0:~}${\it +0.00}$    &{\it 4.08}&{\it 0.40}&{\it \dots}&{\it \dots}&{\it 28.8} \vspace{0.05cm}\\  
0:~$-0.05$          &   3.69   &   0.46   &   \dots   &   \dots   &   33.0    \vspace{0.25cm}\\  
I:~$+0.10$          &   0.45   &   0.81   &    13.4   &   \dots   &   9.77    \vspace{0.05cm}\\  
I:~$+0.05$          &   0.44   &   0.87   &    13.5   &   \dots   &   10.2    \vspace{0.05cm}\\  
{\it I:~}${\it +0.00}$    &{\it 0.42}&{\it 0.93}&{\it 13.5} &{\it \dots}&{\it 11.2} \vspace{0.05cm}\\  
I:~$-0.05$          &   0.40   &   0.99   &    13.6   &   \dots   &   12.5    \vspace{0.25cm}\\  
II:~$+0.10$         &   10.1   &   0.17   &   \dots   &   56.6    &    4.54   \vspace{0.05cm}\\  
II:~$+0.05$         &   9.95   &   0.20   &   \dots   &   64.8    &    4.27   \vspace{0.05cm}\\  
{\it II:~}${\it +0.00}$   &{\it 9.76}&{\it 0.23}&{\it \dots}&{\it 73.1} &{\it 4.73} \vspace{0.05cm}\\  
II:~$-0.05$         &   9.56   &   0.27   &   \dots   &   81.5    &    5.79   \vspace{0.25cm}\\  
III:~$+0.10$        &   0.31   &   0.85   &   19.1    &   30.0    &   1.32    \vspace{0.05cm}\\  
III:~$+0.05$        &   0.42   &   0.83   &   19.0    &   34.8    &   1.16    \vspace{0.05cm}\\  
{\it III:~}${\it +0.00}$  &{\it 0.48}&{\it 0.86}&{\it 18.8} &{\it 38.0} &{\it 1.47} \vspace{0.05cm}\\  
III:~$-0.05$        &   0.50   &   0.90   &   18.7    &   40.0    &   2.18    \vspace{0.10cm}\\  
\hline
\end{tabular}
\note{\tiny Models 0, I, II, and III for $L=4$~kpc, where HEAO-3 data are biased using the formula
$(E_{\rm k/n}/E_c)^{x}$ for $E_{\rm k/n}>E_c$, with $E_c=$~2.9 GeV/n and
$x=+0.1,\,+0.05,\,0.,\,-0.05$.\vspace{-0.25cm}}
\end{table}
However, Model~III (convection and reacceleration), which is statistically preferred over
Model~II, still points to uncomfortably high values for $\delta$. The only way out would be
to have a non-systematic energy effect in the data. For instance, the result on the fit to
the recently published AMS-01 data \citep{2010ICRC0182} leads to $\delta\approx 0.5$ (see Paper~II).

Finally, for completeness, we checked that the effect of the solar modulation parameters
(we took $\Delta\phi \pm50$~MV for the HEAO-3 data) was negligible on all the derived
transport parameters.

\section{Summary and discussion\label{sec:conclusion}}

The guiding questions for this paper were the following. First, do uncertainties
in the input ingredients lead to systematic uncertainties ({\em SystUnc}) on the derived
transport parameters greater than their statistical uncertainties ({\em StatUnc})? Second,
can we reproduce the various values of $\delta$ given in the literature, and which one
should be preferred? We elaborate on the answers below before concluding.

	\subsection{SystUnc and StatUnc: which ones dominate?}

From the analysis of Paper~II, the {\em StatUnc} found for the various transport parameters
typically fall in the range $5\%-10\%$ (see Fig.~\ref{fig:pdfs}). The {\em SystUnc}
generated by each input we varied are summarised below.
\begin{itemize}
	\item A change of $x\%$ in the gas surface density $\Sigma_{\rm ISM}$ translates into a change
	of $x\%$ for all transport parameters but $\delta$, which is unaffected. Such a simple scaling
	can be used for comparing models in the literature, as different
	$\Sigma_{\rm ISM}$ are generally used. Depending on how confident we are in the measurement
	of $\Sigma_{\rm ISM}$, we may conclude that, for this input, the generated {\em SystUnc} 
	is similar to or larger than the {\em StatUnc}.
	\item The transport parameters are very sensitive to the low-energy spectral shape of
	the source parameter $\eta_S$, but not to the source slope $\alpha$. 
	Both parameters can be determined by including data of a primary flux in the fit.
	When doing so, the {\em SystUnc} and {\em StatUnc} are of the same order.
	\item  If we let $\eta_T$|which parameterises the low-energy shape of the diffusion
	coefficient| free in the range $[-2.,1.]$, the {\em SystUnc} completely dominate
	the {\em StatUnc}. Diffusion coefficients with a sharp turn-off at low-energy
	are possible, as shown in Fig.~1 of \citealt{2006ApJ...642..902P} (their dashed
	line corresponds to our $\eta_T=-2$).	
	\item For the cross sections, as seen in Fig.~\ref{fig:pdfs}, the {\em SystUnc} can also 
	be larger than the {\em StatUnc} for some of the transport coefficients. The use of
	other secondary-to-primary ratios, such as Li/C and Be/C, could help to cross-check
	the consistency of each cross-section set, and thus decrease the {\em SystUnc}.
\end{itemize}
These conclusions hold for all classes of models (0, I, II, and III).

	\subsection{Which value of $\delta$ should we trust?}

The two key ingredients for determinaning $\delta$ are i) the low-energy spectral
shape of the diffusion coefficient as parameterised by $\eta_T$ [see Eq.~(\ref{eq:eta_T})], 
and ii) the presence or absence of a wind. These two effects allow $\delta$ to reach values
as low as 0.2 up to 0.9. Such a wide range is also found in \citet{2001ApJ...547..264J},
depending on the model they consider (convection, turbulent diffusion, stochastic reacceleration).

We recover $\delta\approx0.3$ in the pure reacceleration case (no convection),  as in
other propagation codes \citep{1998ApJ...509..212S,2001ApJ...547..264J}. As soon as a
convective wind is included, the diffusion slope is large ($\delta\approx 0.8$), as
obtained in other studies \citep{2001ApJ...547..264J,2001ApJ...555..585M}. The effect of
the low-energy shape of the diffusion coefficient is illustrated by the fact that
standard LBM and standard diffusion models lead to different $\delta$ (more details are
given in Putze, Derome \& Maurin, ICRC 2009). For LBM, which are equivalent to diffusion
models if $K(E)=K_0{\cal R}^\delta$, we recover $\delta\approx 0.5-0.6$
(\citealt{2003ApJS..144..153W}, Paper~I.

\begin{figure}[!t]
\centering
\includegraphics[width = \columnwidth]{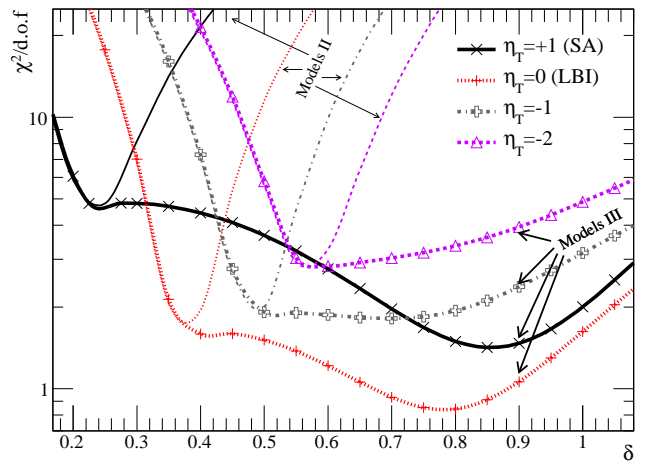}
\caption{Evolution of $\chi^2_{\rm min}$/d.o.f. as a function of $\delta$ for
III (thick lines with symbols) and II (thin lines). The four
cases correspond to different $\eta_T$ (low-energy shape of the diffusion coefficient).}
\label{fig:chi2_Model_IIandIII}
\end{figure}

It is reassuring to see that the results of the various propagation codes used in the
literature are consistent. However, this does not settle the question of which value of
$\delta$ we should trust. In principle, one advantage of our analysis is that it should
clearly point to a preferred value for $\delta$, by a mere comparison of the best
$\chi^2$ values. Indeed, in almost all settings considered, Model~III (convection and
reacceleration) performs better than any other, the second best being Model~II (no
convection). 
This behaviour is illustrated by the $\chi^2_{\rm min}$ dependence as a function of
$\delta$, as shown in Fig.~\ref{fig:chi2_Model_IIandIII}. For a standard diffusion
coefficient (black-solid curve), there is a clear minimum at high $\delta$, and a less
marked {\em local} minimum at low $\delta$. This second minimum matches the $\chi^2_{\rm
min}$ of Model~II (black-dashed curve). The value of the $\chi^2_{\rm min}$ slightly
decreases when $\eta_T$ decreases (red curves), then increases for $\eta_T\lesssim 0$
(grey and violet curves). In the process, the best-fit value for $\delta$ is displaced to
smaller values, the {\em local} minimum at low $\delta$ overtaking and merging with the
high $\delta$ one.

For standard diffusion schemes (SA), the best class of models is Model~III. The main
problem is that whatever ingredients are changed in this model, we are always left with
$\delta\gtrsim 0.6$, and most of the time with even higher values, $\sim 0.9$. Such
high values, always associated to the presence of the constant convective wind, are
difficult to explain. Should we choose to discard this Model~III, we would be left with
the task of explaining why the statistical analysis fails in the context of CR physics. A
way out is provided, as seen in Fig.~\ref{fig:chi2_Model_IIandIII}, if we let free
$\eta_T$ (low-energy shape of the diffusion coefficient). Indeed, the only case where
Model~II (reacceleration only) performs as Model~III is for $\eta_T$ in the
range $-3\lesssim \eta_T\lesssim -1.5$. In such a configuration, the best-fit value for
the convective velocity of Model~III drops to zero, so that the best model only has
reacceleration (i.e. equivalent to Model~II): we term this class of models III/II.
The next best-fit class of models is obtained if we discard reacceleration (Model~I-like)
and again let $\eta_T$ go free. As for the previous case, the best-fit value for the
convective velocity drops to zero: we denote this class I/0. In terms of the best
$\chi^2_{\rm min}$, the standard diffusion Model~II has only the fourth-best $\chi^2$ value.
The best-fit values for these four classes of models are gathered in
Table~\ref{tab:BC_models}, sorted according to their figure of merit.
\begin{table}[!t]
\caption{Best-fit parameters for a few selected configurations.}
\label{tab:BC_models}
\centering
\begin{tabular}{lcccccc} \hline\hline
Model  & $\eta_T$ & $\!\!\!\!\!K_0\times 10^2\!\!\!\!\!$     & $\delta$ &  $V_c$  &  $V_a$  & $\!\!\!\!\chi^2$/d.o.f$\!\!\!\!$   \\
       &          & $\!\!\!\!\!\!\!\!\!$(kpc$^2\,$Myr$^{-1}$)$\!\!\!\!\!\!\!\!\!$ &         &  $\!\!\!\!$(km$\;$s$^{-1}$)$\!\!$  & $\!\!$(km$\;$s$^{-1}$)$\!\!\!\!$   & \\\hline
& \multicolumn{1}{c}{} \vspace{-0.20cm} \\ 
III$^\dagger$    &   SA  & 0.481 & 0.856 & 18.84 & 37.98 & 1.47 \vspace{0.05cm}\\  
III/II$^\star$& -1.3  & 3.161 & 0.512 &   0.  & 45.35 & 2.26 \vspace{0.05cm}\\  
I/0$^\ddagger$    & -2.61 & 2.054 & 0.613 & 0. & \dots & 3.29 \vspace{0.05cm}\\  
II$^\dagger$     &   SA  & 9.753 & 0.234 & \dots & 73.14 & 4.73 \vspace{0.05cm}\\  
\hline
\end{tabular}
{\tiny $\dagger$ Best-fit transport parameters for standard Model~II and~III.}\\
{\tiny $\ddagger$ Best-fit parameters with $\eta_T$ free (no reacceleration).}\\
{\tiny $\star$ Best-fit parameters for fixed $\eta_T$.}
\note{Standard models refer to SA diffusion coefficients (see Table~\ref{tab:Kxx_diff_cases}).
Alternative models show their different values for $\eta_T$, as parameterised from
Eq.~(\ref{eq:eta_T}).\vspace{-0.25cm}}
\end{table}

\begin{figure}[!t]
\centering
\includegraphics[width = \columnwidth]{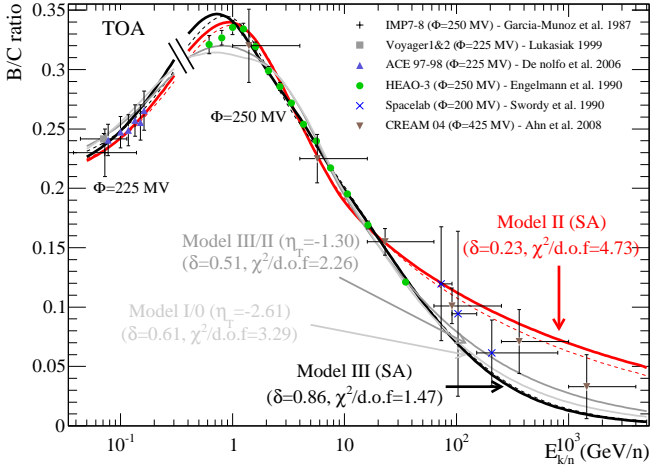}
\caption{Best-fit B/C ratio for standard Model~II (thick red) and~III (thick black) using the {\em reference}
setting (solid and thin dashed lines are for the cross-section sets W03 and GAL09). The special
best-fit Models O/I and III/II are also plotted in thin grey lines (light and dark shade respectively).}
\label{fig:BC_Model_IIandIII}
\end{figure}

The best-fit curves for these four relevant configurations are shown in
Fig.~\ref{fig:BC_Model_IIandIII}. From this graphic view, we can see that Model~II
matches the data best at very-low energy, Model~III at intermediate energies, and Models O/I
and~III/II at high energy. While we found that a systematic bias in HEAO-3 data would not
change the ordering of which is the best model (see Sect.~\ref{sec:biasHEAO}), we see from
Fig.~\ref{fig:BC_Model_IIandIII} that a more complicated pattern in the data may be more
effective in changing which model is best.

To summarise, when a standard scheme for the diffusion coefficient is assumed, Model~III is
always preferred over Model~II. However, the latter seems to agree more with
theoretical expectations for $\delta$. Model~II is also more consistent with the measured
propagated slope $\gamma=\alpha+\delta$. Because of this oddity, we may not be able to give a
definite answer about the value of $\delta$. This is further complicated when $\eta_T$ is left
free; in that case, best models are always without convection, and even the pure diffusive
model is redeemed (although both have difficulty reproducing the B/C peak at GeV/n
energies). We see that many uncertainties show up at GeV/n energies. As also illustrated by
the various predictions for various $\delta$ in Fig~\ref{fig:BC_Model_IIandIII}, the higher the
energy, the closer we can expect to reach the purely diffusive regime. As a result, high-energy B/C
data are desired to unambiguously pinpoint the value of $\delta$.

	\subsection{Conclusion}

In the past years, we have promoted and used a model favouring both convection and reacceleration
\citep[e.g.,][and subsequent studies]{2001ApJ...555..585M,2002A&A...394.1039M} from statistical
criteria. The main and known problem of this model lies in its uncomfortably high value for
$\delta$ ($\delta\sim 0.8$). Such a model is also preferred in the present study. In addition,  we
found that the high value for $\delta$ is extremely resilient to any change in the setting, which
leaves us with several alternatives: assume that there are complicated biases in the data
that conspire to give high $\delta$ in such models, that this high value of $\delta$ is
real (in that case it needs to be explained theoretically), or that any model with convection
should be excluded (which contradicts the fact that winds are observed in many galaxies).
Even if we adopt the last alternative, no firm conclusions can be drawn on the value of $\delta$. Indeed, if the
statistical analysis is relaxed, a large category of models are redeemed, attaining any value for
$\delta$ between 0.3 and 0.9. These models may be purely diffusive, with convection and/or
reacceleration, and are very sensitive to the shape of the low-energy diffusion coefficient (which
is not prescribed theoretically for the moment). Data at higher energy are needed to solve this
question. More constraints can also be obtained by combining several secondary-to-primary ratios
\citep[e.g.,][]{1997AdSpR..19..755W,1997SSRv...81..107W}. This is left for a later study.

This study has limitations. For instance, we only varied the source and diffusion parameters according to
simple parameterisations. More complicated dependences could have been inspected. However, it
is worth recalling that it may be dangerous to introduce too many {\em ad hoc} prescriptions,
because the statistical meaning|already unclear when comparing the different classes of models|becomes less and
less obvious as the number of parameters and models tested increase. In the framework of
homogeneous and isotropic diffusion coefficients, a maybe more important issue is the question
of the Galactic wind. A constant wind was chosen because of the simplicity of the solutions (of the
corresponding diffusion equation). On the one hand, Galactic winds are ubiquitous. On the other, as shown
in this study, constant wind cannot accommodate a realistic slope of the diffusion coefficient.
A linear wind may provide different results. We are implementing a numerical solver in
our propagation code to inspect this issue. Another possibility that cannot be ruled out is that
the HEAO-3 data suffer from non-trivial systematics.

Finally, the starting point of the paper was a comparison between systematic uncertainties
(generated by uncertainties in the input ingredients) and statistical uncertainties for the values
of the transport parameters. That the former can be larger than the latter shows that
many efforts are still needed in CR physics, especially for the production cross-sections, before one
can take full advantage of any statistical analysis, as performed in Papers~I and~II with an MCMC
technique. Some of these issues may be resolved as new data on cosmic-ray nuclei are being released
({\sc cream}, {\sc pamela}, {\sc tracer}).

\begin{acknowledgements}
We thank W.~R.~Webber for providing us with the nuclear production cross-sections discussed
in Webber et al. (2003) and for useful suggestions. We also thank C. Combet for a careful
reading of the paper. A.~P. is grateful for financial support from the Swedish Research
Council (VR) through the Oskar Klein Centre. We acknowledge the support of the French ANR
(grant ANR-06-CREAM).
\end{acknowledgements}

\bibliographystyle{aa}
\bibliography{systematics.bib}

\begin{thebibliography}{27}
\expandafter\ifx\csname natexlab\endcsname\relax\def\natexlab#1{#1}\fi

\bibitem[{{Ahn} {et~al.}(2008){Ahn}, {Allison}, {Bagliesi}, {Beatty},
  {Bigongiari}, {Boyle}, {Brandt}, {Childers}, {Conklin}, {Coutu}, {Duvernois},
  {Ganel}, {Han}, {Hyun}, {Jeon}, {Kim}, {Lee}, {Lee}, {Lutz}, {Maestro},
  {Malinin}, {Marrocchesi}, {Minnick}, {Mognet}, {Nam}, {Nutter}, {Park},
  {Park}, {Seo}, {Sina}, {Swordy}, {Wakely}, {Wu}, {Yang}, {Yoon}, {Zei}, \&
  {Zinn}}]{2008APh....30..133A}
{Ahn}, H.~S., {Allison}, P.~S., {Bagliesi}, M.~G., {et~al.} 2008, Astroparticle
  Physics, 30, 133

\bibitem[{{Ave} {et~al.}(2008){Ave}, {Boyle}, {Gahbauer}, {H{\"o}ppner},
  {H{\"o}randel}, {Ichimura}, {M{\"u}ller}, \&
  {Romero-Wolf}}]{2008ApJ...678..262A}
{Ave}, M., {Boyle}, P.~J., {Gahbauer}, F., {et~al.} 2008, \apj, 678, 262

\bibitem[{{de Nolfo} {et~al.}(2006){de Nolfo}, {Moskalenko}, {Binns},
  {Christian}, {Cummings}, {Davis}, {George}, {Hink}, {Israel}, {Leske},
  {Lijowski}, {Mewaldt}, {Stone}, {Strong}, {von Rosenvinge}, {Wiedenbeck}, \&
  {Yanasak}}]{2006AdSpR..38.1558D}
{de Nolfo}, G.~A., {Moskalenko}, I.~V., {Binns}, W.~R., {et~al.} 2006, Advances
  in Space Research, 38, 1558

\bibitem[{{Donato} {et~al.}(2009){Donato}, {Maurin}, {Brun}, {Delahaye}, \&
  {Salati}}]{2009PhRvL.102g1301D}
{Donato}, F., {Maurin}, D., {Brun}, P., {Delahaye}, T., \& {Salati}, P. 2009,
  Phys. Rev. Lett., 102, 071301

\bibitem[{{Engelmann} {et~al.}(1990){Engelmann}, {Ferrando}, {Soutoul},
  {Goret}, \& {Juliusson}}]{1990A&A...233...96E}
{Engelmann}, J.~J., {Ferrando}, P., {Soutoul}, A., {Goret}, P., \& {Juliusson},
  E. 1990, \aap, 233, 96

\bibitem[{{Garcia-Munoz} {et~al.}(1987){Garcia-Munoz}, {Simpson}, {Guzik},
  {Wefel}, \& {Margolis}}]{1987ApJS...64..269G}
{Garcia-Munoz}, M., {Simpson}, J.~A., {Guzik}, T.~G., {Wefel}, J.~P., \&
  {Margolis}, S.~H. 1987, \apj Supp. Series, 64, 269

\bibitem[{{Jones} {et~al.}(2001){Jones}, {Lukasiak}, {Ptuskin}, \&
  {Webber}}]{2001ApJ...547..264J}
{Jones}, F.~C., {Lukasiak}, A., {Ptuskin}, V., \& {Webber}, W. 2001, \apj, 547,
  264

\bibitem[{{Lionetto} {et~al.}(2005){Lionetto}, {Morselli}, \&
  {Zdravkovic}}]{2005JCAP...09..010L}
{Lionetto}, A.~M., {Morselli}, A., \& {Zdravkovic}, V. 2005, JCAP, 9, 10

\bibitem[{{Lukasiak} {et~al.}(1999){Lukasiak}, {McDonald}, \&
  {Webber}}]{1999ICRC....3...41L}
{Lukasiak}, A., {McDonald}, F.~B., \& {Webber}, W.~R. 1999, 3, 41

\bibitem[{{Maurin} {et~al.}(2001){Maurin}, {Donato}, {Taillet}, \&
  {Salati}}]{2001ApJ...555..585M}
{Maurin}, D., {Donato}, F., {Taillet}, R., \& {Salati}, P. 2001, \apj, 555, 585

\bibitem[{{Maurin} {et~al.}(2006){Maurin}, {Taillet}, \&
  {Combet}}]{2006astro.ph.12714M}
{Maurin}, D., {Taillet}, R., \& {Combet}, C. 2006, astro-ph/0612714

\bibitem[{{Maurin} {et~al.}(2002){Maurin}, {Taillet}, \&
  {Donato}}]{2002A&A...394.1039M}
{Maurin}, D., {Taillet}, R., \& {Donato}, F. 2002, A\&A, 394, 1039

\bibitem[{{Osborne} \& {Ptuskin}(1988)}]{1988SvAL...14..132O}
{Osborne}, J.~L. \& {Ptuskin}, V.~S. 1988, Sov. Astron. Lett., 14, 132

\bibitem[{{Ptuskin} {et~al.}(2006){Ptuskin}, {Moskalenko}, {Jones}, {Strong},
  \& {Zirakashvili}}]{2006ApJ...642..902P}
{Ptuskin}, V.~S., {Moskalenko}, I.~V., {Jones}, F.~C., {Strong}, A.~W., \&
  {Zirakashvili}, V.~N. 2006, \apj, 642, 902

\bibitem[{{Putze} {et~al.}(2010){Putze}, {Derome}, \&
  {Maurin}}]{2009A&A...xxx..xxxP}
{Putze}, A., {Derome}, L., \& {Maurin}, D. 2010, \aap\ accepted (Paper II)

\bibitem[{{Putze} {et~al.}(2009){Putze}, {Derome}, {Maurin}, {Perotto}, \&
  {Taillet}}]{2009A&A...497..991P}
{Putze}, A., {Derome}, L., {Maurin}, D., {Perotto}, L., \& {Taillet}, R. 2009,
  \aap, 497, 991 (Paper I)

\bibitem[{{Schlickeiser}(2002)}]{2002cra..book.....S}
{Schlickeiser}, R. 2002, {Cosmic Ray Astrophysics}, ed. R.~{Schlickeiser}

\bibitem[{{Seo} \& {Ptuskin}(1994)}]{1994ApJ...431..705S}
{Seo}, E.~S. \& {Ptuskin}, V.~S. 1994, \apj, 431, 705

\bibitem[{{Strong} \& {Moskalenko}(1998)}]{1998ApJ...509..212S}
{Strong}, A.~W. \& {Moskalenko}, I.~V. 1998, \apj, 509, 212

\bibitem[{{Swordy} {et~al.}(1990){Swordy}, {Mueller}, {Meyer}, {L'Heureux}, \&
  {Grunsfeld}}]{1990ApJ...349..625S}
{Swordy}, S.~P., {Mueller}, D., {Meyer}, P., {L'Heureux}, J., \& {Grunsfeld},
  J.~M. 1990, \apj, 349, 625

\bibitem[{{Taillet} \& {Maurin}(2003)}]{2003A&A...402..971T}
{Taillet}, R. \& {Maurin}, D. 2003, \aap, 402, 971

\bibitem[{{Tomassetti} \& {AMS-01 Collaboration}(2009)}]{2010ICRC0182}
{Tomassetti}, N. \& {AMS-01 Collaboration}. 2009, in International Cosmic Ray
  Conference, International Cosmic Ray Conference, 1--4

\bibitem[{{Webber}(1997{\natexlab{a}})}]{1997AdSpR..19..755W}
{Webber}, W.~R. 1997{\natexlab{a}}, Adv. Space Res., 19, 755

\bibitem[{{Webber}(1997{\natexlab{b}})}]{1997SSRv...81..107W}
{Webber}, W.~R. 1997{\natexlab{b}}, Space Sci. Rev., 81, 107

\bibitem[{{Webber} {et~al.}(1998){Webber}, {Kish}, {Rockstroh}, {Cassagnou},
  {Legrain}, {Soutoul}, {Testard}, \& {Tull}}]{1998ApJ...508..940W}
{Webber}, W.~R., {Kish}, J.~C., {Rockstroh}, J.~M., {et~al.} 1998, \apj, 508,
  940

\bibitem[{{Webber} {et~al.}(1990){Webber}, {Kish}, \&
  {Schrier}}]{1990PhRvC..41..566W}
{Webber}, W.~R., {Kish}, J.~C., \& {Schrier}, D.~A. 1990, \prc, 41, 566

\bibitem[{{Webber} {et~al.}(2003){Webber}, {Soutoul}, {Kish}, \&
  {Rockstroh}}]{2003ApJS..144..153W}
{Webber}, W.~R., {Soutoul}, A., {Kish}, J.~C., \& {Rockstroh}, J.~M. 2003,
  \apjs, 144, 153

\end{thebibliography}
\end{document}